\documentclass[fleqn,usenatbib]{mnras}

\usepackage[T1]{fontenc}
\usepackage{color}
\usepackage{float}

\DeclareRobustCommand{\VAN}[3]{#2}
\let\VANthebibliography\thebibliography
\def\thebibliography{\DeclareRobustCommand{\VAN}[3]{##3}\VANthebibliography}

\usepackage{graphicx}	
\usepackage{amsmath}	
\usepackage{amssymb}	
\usepackage{caption}
\usepackage{subcaption}
\usepackage{newtxtext,newtxmath}

\newcommand{\red}[1]{\textcolor{black}{#1}}

\title[TNG x DESI]{Illustrating galaxy-halo connection in the DESI era with \textsc{IllustrisTNG}}

\author[S. Yuan et al.]{
Sihan Yuan,$^{1,2}$\thanks{E-mail: sihan.yuan@cfa.harvard.edu}
Boryana Hadzhiyska, $^{1}$
Sownak Bose, $^{1,3}$
and Daniel J. Eisenstein $^{1}$
\\
$^{1}$Center for Astrophysics | Harvard \& Smithsonian, 60 Garden St., Cambridge, MA 02138, USA\\
$^{2}$Kavli Institute for Particle Astrophysics and Cosmology, Stanford University, Stanford, CA 94305, USA\\
$^{3}$Institute for Computational Cosmology, Department of Physics, Durham University, Durham DH1 3LE, UK
}

\date{Accepted XXX. Received YYY; in original form ZZZ}

\pubyear{2015}

\begin{document}
\label{firstpage}
\pagerange{\pageref{firstpage}--\pageref{lastpage}}
\maketitle

\begin{abstract}
We employ the hydrodynamical simulation \textsc{IllustrisTNG} to inform the galaxy-halo connection of the Luminous Red Galaxy (LRG) and Emission Line Galaxy (ELG) samples of the Dark Energy Spectroscopic Instrument (DESI) survey at redshift $z \sim 0.8$. Specifically, we model the galaxy colors of \textsc{IllustrisTNG} and apply sliding DESI color-magnitude cuts, matching the DESI target densities. We study the halo occupation distribution model (HOD) of the selected samples by matching them to their corresponding dark matter halos in the \textsc{IllustrisTNG} dark matter run. We find the HOD of both the LRG and ELG samples to be consistent with their respective baseline models, but also we find important deviations from common assumptions about the satellite distribution, velocity bias, and galaxy secondary biases. We identify strong evidence for concentration-based and environment-based occupational variance in both samples, an effect known as ``galaxy assembly bias''. The central and satellite galaxies have distinct dependencies on secondary halo properties, showing that centrals and satellites have distinct evolutionary trajectories and should be modelled separately. These results serve to inform the necessary complexities in modeling galaxy-halo connection for DESI analyses and also prepare for building high-fidelity mock galaxies. Finally, we present a shuffling-based clustering analysis that reveals a 10--15$\%$ excess in the LRG clustering of modest statistical significance due to secondary galaxy biases. We also find a similar excess signature for the ELGs, but with much lower statistical significance. When a larger hydrodynamical simulation volume becomes available, we expect our analysis pipeline to pinpoint the exact sources of such excess clustering signatures. 
\end{abstract}




\section{Introduction}
In the standard framework of structure formation in a $\Lambda$CDM universe, galaxies are predicted to form and evolve in dark matter halos \citep{1978White}.
To extract cosmological information and understand galaxy formation from  observed galaxy clustering statistics, it is critical to correctly model the connection between galaxies and their underlying dark matter halos. The most popular and efficient model of the galaxy-halo connection for cosmological studies is the Halo Occupation Distribution model \citep[HOD; e.g.][]{2000Peacock, 2001Scoccimarro, 2002Berlind, 2005Zheng, 2007bZheng}.
The HOD model provides a simple empirical relation between halo mass and the number of galaxies it hosts, which is expressed as the probability distribution $P(N_g|M_h)$ that a halo of mass $M_h$ hosts $N_g$ galaxies satisfying some selection criteria.
The HOD model is thus particularly well suited to study galaxy clustering \citep[e.g.][]{2005Zheng, 2011Zehavi, 2015aGuo, 2021Yuan}, since the HOD parameters can be tuned so as to reproduce a set of observables such as the two-point correlation function and the galaxy number density, 

The HOD model is also markedly flexible as one can easily introduce extensions to incorporate additional physics that might affect galaxy occupation \citep[e.g.][]{, 2016Hearin, 2018Yuan, 2020Xu}. This is particularly important as we attain high precision clustering measurements at nonlinear scales, where additional non-linear biases and halo-scale physics need to be modeled to accurately reproduce the observed clustering. For example, in \citet{2021Yuan}, we found that the addition of secondary dependencies on halo concentration and environment in the HOD model significantly improves its ability to predict the full-shape galaxy clustering and galaxy-galaxy lensing on small-scales. Another well known extension to the HOD model is velocity bias, which \citet{2015aGuo} and \citet{2021bYuan} found to be a critical model ingredient to accurately reproduce the observed redshift-space galaxy clustering. 

Refining and testing these extensions to the HOD model is becoming even more important with the new generation of galaxy surveys. The Dark Energy Spectroscopic Instrument \citep[DESI;][]{2016DESI} is an on-going flagship spectroscopic survey that will eventually precisely measure the 3D positions of more than 30 million objects over a 14,000 deg$^2$ survey footprint, mapping Luminous Red Galaxies (LRGs) and Emission Line Galaxies (ELGs) up to redshift 1.6. \red{This is significantly deeper than and approximately 10 times the effective volume of the current state-of-art Baryon Oscillation Spectroscopic Survey \citep[BOSS;][]{2012Bolton, 2013Dawson} and the extended Baryon Oscillation Spectroscopic Survey \citep[eBOSS;][]{2016Dawson, 2017Blanton} dataset. We note that eBOSS specifically can be thought of as a predecessor to DESI, mapping a similar set of galaxy tracers in a comparable redshift range albeit with a much smaller footprint. We will present several key comparisons with eBOSS results in this analysis.} On the large scales, DESI will leverage the baryon acoustic oscillations (BAO) and redshift-space distortions (RSD) measurements to place stringent constraints on models of dark energy, modified gravity and inflation, as well as the neutrino mass. 
\red{However, DESI and other upcoming surveys are designed in a way that sees most gains in statistical power on modestly nonlinear scales ($\sim10h^{-1}$Mpc) down to small scales ($< 1h^{-1}$Mpc). The clustering on small scales is expected to be feature-rich and contain a vast amount of information on cosmology and galaxy formation physics \citep[e.g.][]{2019Zhai, 2021Lange}.} However, revealing this information requires a robust model of galaxy-dark matter connection, with extensions to describe secondary biases and various halo-scale physics.  

While it is important to extend the HOD model to make it sufficiently flexible in modeling clustering on non-linear scales, it is also essential to validate each extension before applying them to data to avoid overfitting and to ensure that the extensions are physically motivated. Full-physics hydrodynamical simulations provide the ideal avenue for validation by simulating the galaxy formation process simultaneously with dark matter evolution \citep[e.g.][]{2020Vogelsberger, 2018Hopkins, 2014bVogelsberger, 201Schaye, 2003Abadi}. With sufficient volume, one can directly test and calibrate any galaxy-dark matter connection model in a hydrodynamical simulation, assuming that simulation closely mimics reality \citep[e.g.][]{2021Delgado, 2021bHadzhiyska, 2020Hadzhiyska}. However, the ability to simulate the details of galaxy formation comes at a steep computational cost. Specifically, the state-of-the-art hydrodynamical simulations achieve high fidelity by incorporating various baryonic effects such as stellar wind, supernova feedback, gas cooling, and black hole feedback. Because of such high complexity, full-physics hydrodynamical simulations have only recently reached sizes of a few hundred megaparsec \citep[][]{2016Chaves, 2018Springel}. While such volume is still not enough for cosmological studies, it begins to offer sufficient statistical power to constrain galaxy-halo connection models and their small-scale clustering signatures. 

In this paper, we leverage the state-of-the-art \textsc{IllustrisTNG} hydrodynamical simulation \citep[e.g.][]{2018Pillepich,2018Springel,2018Nelson} to calibrate the HOD model and its necessary extensions for DESI galaxy samples. Specifically, we apply DESI LRG and ELG selection to \textsc{IllustrisTNG} galaxies and directly study their connection to the underlying dark matter halos. We validate a series of extensions such as satellite radial bias, velocity bias, and concentration-based and environment-based secondary bias. This study produces the most detailed galaxy-halo connection model for DESI galaxy samples, which not only aids in the creation of realistic mock catalogs on much large dark matter only simulations, but also paves the way for analyzing the upcoming DESI full-shape clustering measurements. The findings of this analysis also serve to inform galaxy-halo connection modeling needs in current and upcoming cosmological surveys, such as the Dark Energy Spectroscopic Instrument \citep[DESI;][]{2016DESI}, the Subaru Prime Focus Spectrograph \citep[PFS;][]{2014Takada}, the ESA \textit{Euclid} satellite mission \citep[][]{2011Laureijs}, and the NASA \textit{Roman Space Telescope} \citep[WMAP;][]{2013Spergel}. 

This paper is organized as follows. In Section~\ref{sec:method}, we describe the selection of DESI LRG and ELG mocks from the \textsc{IllustrisTNG} simulation box. In Section~\ref{sec:results}, we present the baseline HOD of the two samples, and examine the need for various HOD extensions for these samples. In Section~\ref{sec:discuss}, we compare our results to previous studies and put this work in the broader context of cosmological analysis through numerical simulations. Finally, we conclude in Section~\ref{sec:conclusion}.

\section{Methodology}
\label{sec:method}
In this section, we introduce the simulations we use, and how we extract galaxy samples and their dark matter halo counterparts from such simulations. 
 
\subsection{\textsc{IllustrisTNG}}
Our galaxy populations are drawn from the state-of-the-art hydrodynamical simulation suite \textsc{IllustrisTNG} \citep[][]{2018Pillepich,2018Marinacci,2018Naiman,
2018Springel,2019Nelson,2018Nelson,2019Pillepich,2019bNelson}. 
\textsc{IllustrisTNG} is a suite of cosmological magneto-hydrodynamic simulations,
which were carried out using the \textsc{AREPO}
code \citep{2010Springel} with cosmological parameters consistent with the
\textit{Planck 2015} analysis \citep{2016Planck}.
These simulations feature a series of improvements
compared with their predecessor, \textsc{Illustris}, such as
improved kinetic AGN feedback and galactic wind
models, as well as the inclusion of magnetic fields.

In particular, we utilize the \textsc{IllustrisTNG}-300-1 box, the largest high-resolution hydrodynamical simulation from the suite. 
The size of its periodic box is 205$h^{-1}$Mpc with 2500$^3$ DM particles
and 2500$^3$ gas cells, implying a DM particle mass of $3.98 \times 10^7 \ h^{-1}\rm{M_\odot}$ and
baryonic mass of $7.44 \times 10^6 \ h^{-1}\rm{M_\odot}$. 
We also use the dark-matter-only (DMO) counterpart of the \textsc{IllustrisTNG}-300-1 box, \textsc{IllustrisTNG}-300-Dark, which was evolved with the same initial conditions and the same number of dark matter particles ($2500^3$), each with particle mass of $4.73\times 10^7 h^{-1}M_\odot$.

The haloes (groups) in \textsc{IllustrisTNG}-300-Dark are found with a standard 
friends-of-friends (FoF) algorithm with linking length $b=0.2$ (in units of the mean interparticle spacing)
run on the dark matter particles, while the subhaloes are identified 
using the SUBFIND algorithm \citep{2000Springel}, which detects 
substructure within the groups and defines locally overdense, self-bound particle groups.
For this paper, we analyse the simulations at redshift $z = 0.8$. 

The key analysis in this paper is performed by selecting galaxies from the \textsc{IllustrisTNG}-300-1 box and matching them to their halo counterparts in \textsc{IllustrisTNG}-300-Dark. We describe this process in detail in the following subsections.

\subsection{Generating galaxy colors}
To select DESI-like galaxies in \textsc{IllustrisTNG}, we first need to generate colors for each stellar subhalo in \textsc{IllustrisTNG} so that we can apply DESI color-magnitude selection cuts.
We follow the same procedure as \citet{2021bHadzhiyska} in using a stellar population synthesis and dust model to generate mock galaxy colors. Specifically, we use the Flexible Stellar Population Synthesis code \citep[FSPS,][]{2010Conroy,2010bConroy}. We adopt the MILES stellar library \citep{2015Vazdekis} and the MIST isochrones \citep{2016Choi}. We measure the star-formation history in the simulation from all the stellar particles in a subhalo within 30 kpc of its center. We split the star-formation history of each galaxy into a young (stellar ages $<30$ Myr) and old (stellar ages $>30$ Myr) component. \red{We justify the choice of 30 Myr by noting that, as shown in e.g., Fig. 2 of \citet{2017ApJ...840...44B}, at time scales longer than 30 Myr, there are very few ionising photons.} We use the young SFH component to predict the nebular continuum emission and emission lines, assuming the measured gas-phase metallicity from the simulation and $-$1.4 for \red{the log gas ionization paremeter, \texttt{logu}, defined in Eq.~2 of \citet{2017ApJ...840...44B} and relevant only for the nebular continuum emission}. We feed the old SFH component along with the mass-weighted stellar metallicity history to FSPS in order to predict the stellar continuum emission. For the galaxies studied here, the latter component dominates the flux in the DESI photometric bands. 

\red{There are different ways of how to model dust attenuation in the simulation \citep[e.g.,][]{2018Nelson,2020Vogelsberger}}. Here, we use an empirical approach by basing our attenuation prescription on recent observational measurements. Specifically, we assume that the absorption optical depth follows:

\begin{equation}
    \tau_{\rm v} = \gamma \left(\frac{Z_{\rm gas}}{Z_{\odot}}\right)^{\alpha} \tilde{\Sigma}_{\star}^{\beta},
\end{equation}

\noindent
where $Z_{\rm gas}$ is the gas-phase metallicity and $\tilde{\Sigma}$ is the normalized stellar mass density ($\tilde{\Sigma}=\Sigma_{\star}/\langle\Sigma_{\star}\rangle$ with $\Sigma_{\star}=M_{\star}/(\pi r_{\rm e}^2$), where $r_{\rm e}$ is the half-mass radius). Both quantities are obtained directly from the simulations. The parameters $\alpha$, $\beta$ and $\gamma$ have been roughly tuned to reproduce observed $z\sim0$ scaling relations between $\tau_{\rm v}$, SFR and $M_{\star}$ by \cite{2018Salim}, \red{which is based on GALEX, SDSS and WISE photometry \citep[][]{2016Beitia, 2016Lang}.} Specifically, we find $\alpha$, $\beta$ and $\gamma$ to be $-0.6$, $0.2$, $0.4$, respectively. We also vary the additional dust attenuation toward younger stars (\texttt{dust1} in FSPS) and the dust index (shape of the dust attenuation law), and find that values close to the standard values within FSPS reproduces well the observed colour distribution at the redshifts of interest (shown in Section~2.4.3 of \citealt{2021bHadzhiyska}). We emphasize that the dust model parameters are only roughly tuned to observations and we did not formally optimize the dust model parameters.

A more detailed description of the galaxy color models can be found in \citet{2021bHadzhiyska}.

\subsection{Selecting DESI-like galaxies}

\begin{figure}
    \centering
    \hspace*{-0.2cm}
    \includegraphics[width = 3.3in]{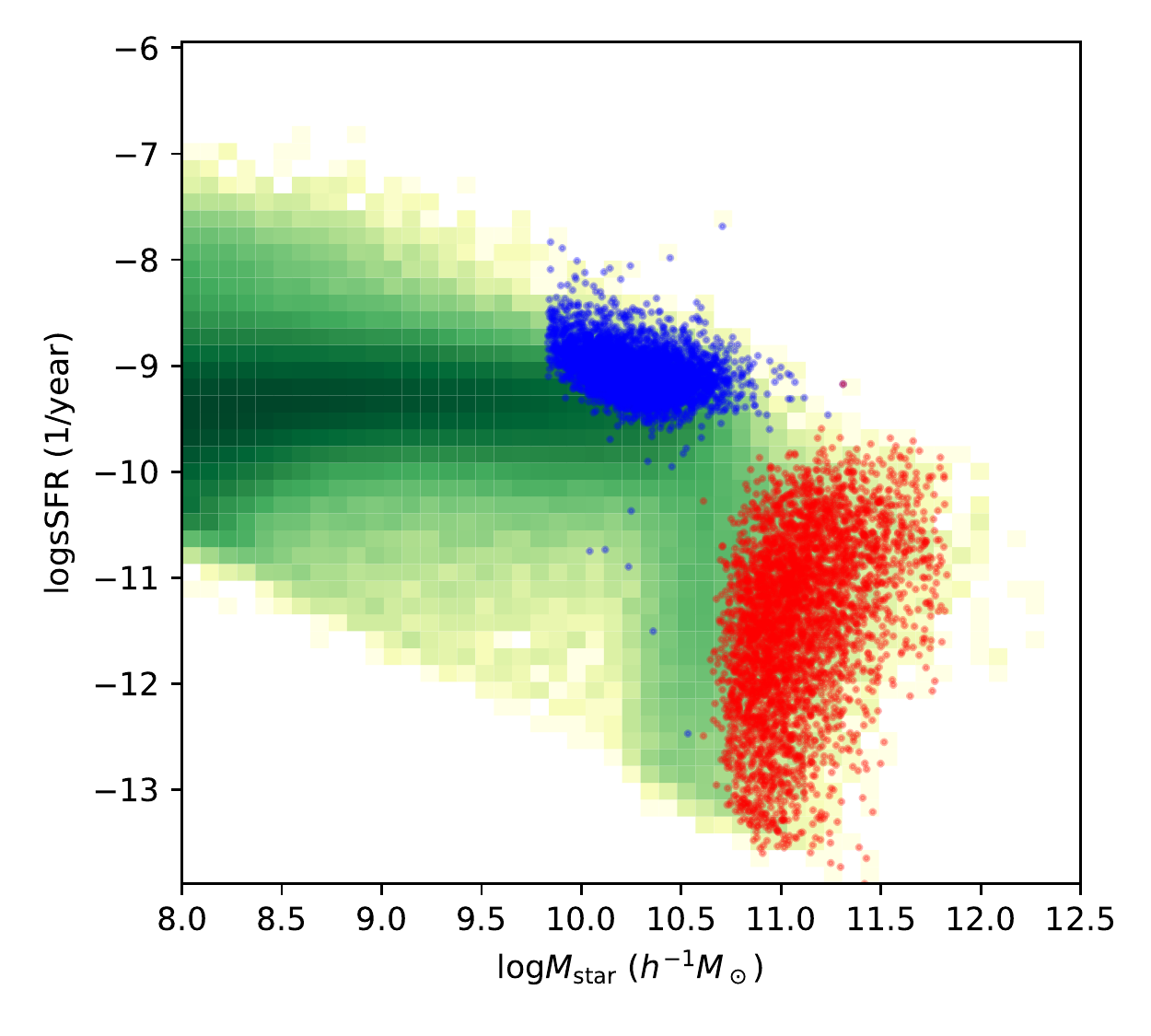}
    \vspace{-0.3cm}
    \caption{The distribution of the DESI LRG/ELG mock sample on the stellar mass vs. specific star formation rate plane at $z = 0.8$. \red{The stellar mass and star formation data are taken from the \textsc{IllustrisTNG} subhalo catalogs.} The colored histogram shows the distribution of all galaxies in the simulation, above $M_{\mathrm{star}} > 10^8 M_\odot$, showing two distinct populations. The red points showcase the distribution of our LRG mock sample, unsurprisingly occupying the high mass, low star formation rate corner of the plot. The blue points showcase the ELG mock sample, which appears to be well localized in a lower mass higher star formation rate cluster.}
    \label{fig:lrg_sample}
\end{figure}
Having obtained the galaxy color magnitudes, it is in principle trivial to apply the color cuts from DESI target selection to obtain DESI-like galaxy mocks. However, there is potentially a systematic bias in our color magnitude model, due to differences in how the magnitudes are actually measured on the sky and how they are modeled in simulations. Thus, we apply a floating magnitude correction $\delta m$ to all the modeled galaxy magnitudes. We calibrate this $\delta m$ by approximately matching the actual target number density in DESI. 
\subsubsection{DESI LRGs}
For the LRG mock, we use the sliding color cuts of DESI LRG target selection \citep{2020Zhou}:
\begin{align}
    & (r' - z') > (z' - 16.83) \times 0.45,\\ 
    & \mathrm{AND}\ (r' - z') > (z' -3.80) \times 0.19,
    \label{equ:lrg_cuts}
\end{align}
where $r' = r_{\mathrm{model}} + \delta m$ and $z'= z_{\mathrm{model}} + \delta m$ are the corrected model magnitudes. The target number density of LRGs in DESI in the redshift range $z=0.6-1.05$ is $5\times 10^{-4}h^3$Mpc$^{-3}$. We find that $\delta m = -0.4$ roughly matches the desired target density, resulting in an LRG mock sample of 4608 galaxies. 

Figure~\ref{fig:lrg_sample} showcases the distribution of the LRG mock sample on the stellar mass vs. specific star formation rate plane. The underlying color histogram shows the distribution of all the galaxies (subhalos) in the simulation boxabove $M_{\mathrm{star}} > 10^8 h^{-1}M_\odot$, whereas the red points represent the LRG sample. \red{The star formation rate and the stellar mass information were taken directly from the \textsc{IllustrisTNG} subhalo catalogs. Specifically, the stellar mass of a galaxy/subhalo is defined as the total mass of all member particle/cells which are bound to this subhalo, and the star formation rate is defined as the sum of the individual star formation rates of all gas cells in this subhalo.} We see clearly the existence of two galaxy populations, with the dominant population being young star-forming galaxies with lower stellar mass. The bottom right corner represents a second population of massive, but more quenched galaxies. As we expected, the LRGs (shown in red) occupy the more massive less star-forming end of this population. 

\subsubsection{DESI ELGs}
Similarly for ELG mocks, we adopt the sliding color cuts from DESI target selection \citep[][]{2016DESI, 2020Raichoor}:
\begin{align}
    & 0.3 < (r'-z')< 1.6,\\ 
    & \mathrm{AND}\ (g'-r') < 1.15 \times (r'-z')-0.15,\\
    & \mathrm{AND}\ (g'-r')< -1.2 \times (r'-z')+1.6,
    \label{equ:elg_cuts}
\end{align}
where $g', r', z'$ again symbolize the corrected model magnitudes. The target number density of ELGs in DESI in the redshift range $z = 0.6-1.05$ is $5\times 10^{-4}h^3$Mpc$^{-3}$. The corresponding magnitude correction that approximates this number density is $\delta m = 0.6$, resulting in a sample of 4998 ELGs. \red{We display the mock ELG sample in the stellar mass vs. star formation rate plane by the blue points in Figure~\ref{fig:lrg_sample}. Compared to the mock LRGs, the mock ELGs are well localized in a lower stellar mass higher star formation rate cluster.}

\subsection{Identifying corresponding DMO halos}
To evaluate the HOD of the selected mock galaxies, we also need to identify their dark matter counterparts in the DMO simulation. The existing TNG outputs provide the bijective mapping between most of the subhalos in the full-physics simulation and the DMO simulation. This provides a straightforward avenue to map the full-physics halos and the DMO halos, by matching their most massive subhalos. However, a small subset of full-physics subhalos do not have available DMO counterparts. Thus, for these objects, we manually map them to DMO subhalos by proximity in 3D position and mass. This way, we successfully identify host DMO halos for every DESI-like galaxy selected in this analysis. 

For each halo, we use $M_{200c}$ as its halo mass. Specifically, $M_{200c}$ refers to the mass enclosed in $r_{200c}$, which is radius within which the halo has an overdensity 200 times the critical density of the Universe. We use the default outputs of \textsc{IllustrisTNG} for the halo position and velocity, corresponding to the position of the particle with the minimum gravitational potential energy and the sum of the mass weighted velocities of all particles/cells in the halo, respectively. 

\section{Results}
\label{sec:results}
Having selected the mock LRG and ELG samples and their corresponding halos, we present the key HOD measurements in this section. 
\subsection{Baseline HOD}

\begin{figure}
    \centering
    \hspace*{-0.6cm}
    \includegraphics[width = 3.4in]{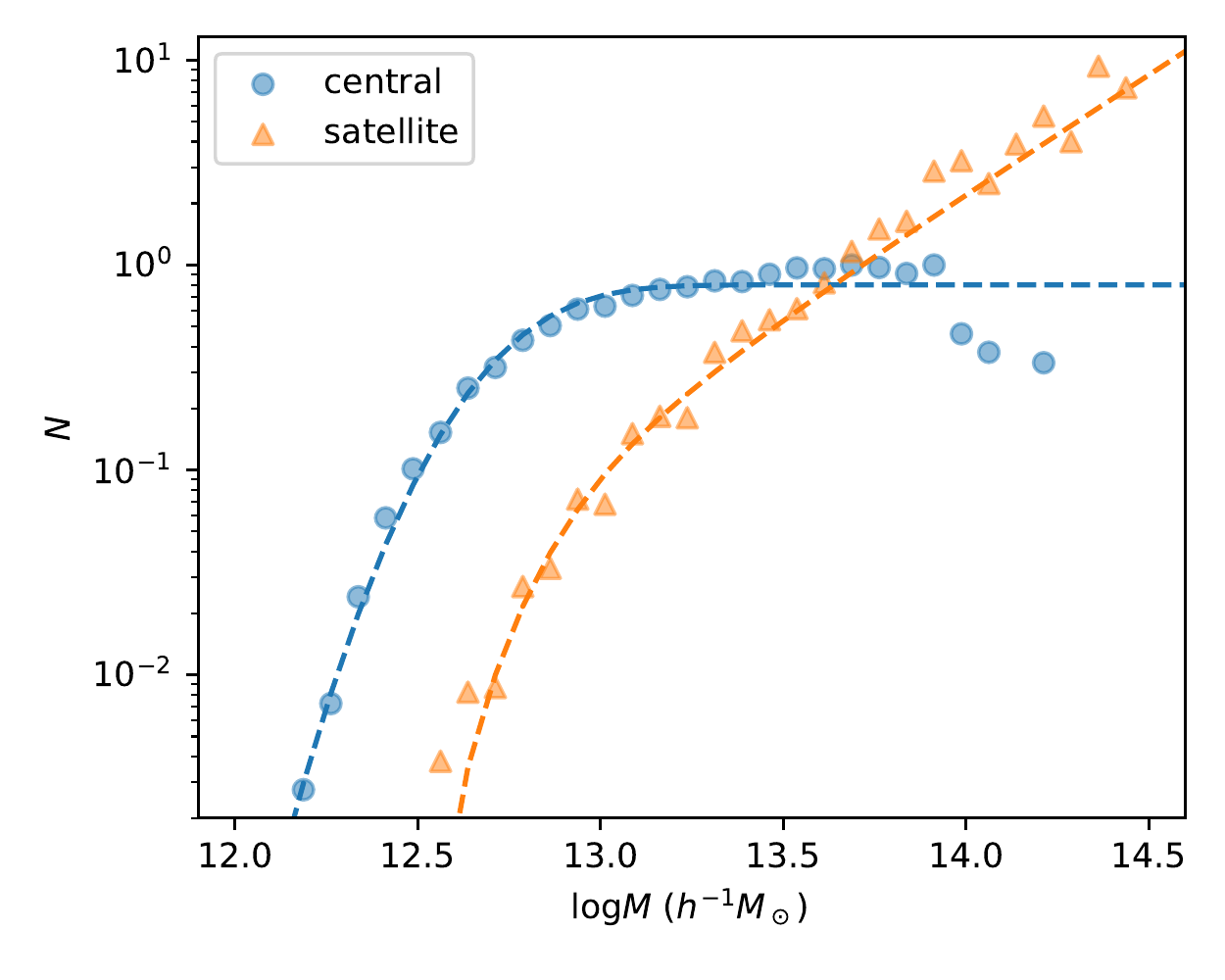}
    \vspace{-0.3cm}
    \caption{The HOD of the DESI-LRG mocks in \textsc{IllustrisTNG}. The dots correspond to the measured HOD of the LRG mocks, whereas the dashed lines correspond to a fiducial baseline model chosen to roughly match the measurements. Blue corresponds to central galaxies, whereas orange corresponds to satellite galaxies. The 5-parameter baseline HOD model plus incompleteness can reproduce the measured mass dependence reasonably well. The dashed lines correspond to $\log M_\mathrm{cut} = 12.7$, $\log M_1 = 13.6$, $\sigma = 0.2$, $\alpha = 1.15$, $\kappa = 0.08$, and incompleteness $f_\mathrm{ic} = 0.8$. }
    \label{fig:lrg_hod}
\end{figure}
\begin{figure}
    \centering
    \hspace*{-0.6cm}
    \includegraphics[width = 3.4in]{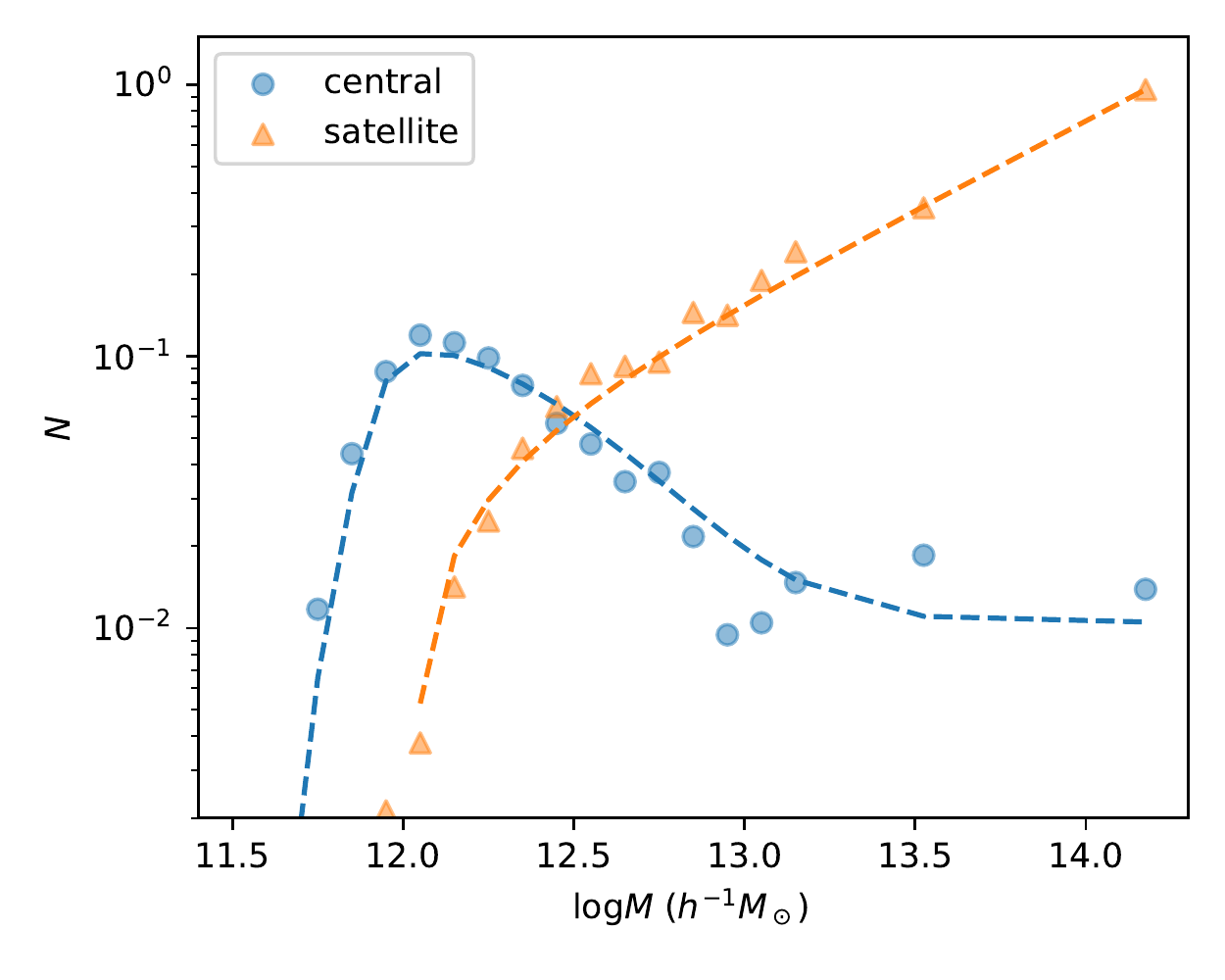}
    \vspace{-0.3cm}
    \caption{The HOD of the DESI-ELG mocks in \textsc{IllustrisTNG}. The dots correspond to the measured HOD of the ELG mocks, whereas the dashed lines correspond to a fiducial baseline model chosen to roughly match the measurements. Blue corresponds to central galaxies, whereas orange corresponds to satellite galaxies. The baseline HOD model can reproduce the measured mass dependence reasonably well. The measured central occupation at $\log M > 13.5$ is noisy due to the small number of ELG host halos in this mass range. The dashed lines correspond to $p_\mathrm{max} = 0.075$, $Q = 95$, $\log M_\mathrm{cut} = 11.9$, $\sigma = 0.5$, $\gamma = 5$, $\log M_1 = 14.2$, $\alpha = 0.65$, and $\kappa = 1.35$.}
    \label{fig:elg_hod}
\end{figure}
In this subsection, we examine how the TNG mocks compare to the baseline HOD models. For LRGs, the baseline model refers to the 5-parameter model from \citet{2007bZheng}, which gives the mean expected number of central and satellite galaxies per halo given halo mass:
\begin{align}
    N_{\mathrm{cent}}^{\mathrm{LRG}}(M) & = \frac{f_\mathrm{ic}}{2}\mathrm{erfc} \left[\frac{\log_{10}(M_{\mathrm{cut}}/M)}{\sqrt{2}\sigma}\right], \label{equ:zheng_hod_cent}\\
    N_{\mathrm{sat}}^{\mathrm{LRG}}(M) & = \left[\frac{M-\kappa M_{\mathrm{cut}}}{M_1}\right]^{\alpha}N_{\mathrm{cent}}^{\mathrm{LRG}}(M),
    \label{equ:zheng_hod_sat}
\end{align}
where the five baseline parameters characterizing the model are $M_{\mathrm{cut}}, M_1, \sigma, \alpha, \kappa$. $M_{\mathrm{cut}}$ characterizes the minimum halo mass to host a central galaxy. $M_1$ characterizes the typical halo mass that hosts one satellite galaxy. $\sigma$ describes the steepness of the transition from 0 to 1 in the number of central galaxies. $\alpha$ is the power law index on the number of satellite galaxies. $\kappa M_\mathrm{cut}$ gives the minimum halo mass to host a satellite galaxy. In addition to the baseline parameters, we have also added the incompleteness parameter $f_\mathrm{ic}$, which is introduced to modulate the overall completeness and density of the sample. By definition, $0 < f_\mathrm{ic} \leq 1$, with $f_\mathrm{ic} = 1$ corresponding to a complete sample. 

We have also added a conformity term $N_{\mathrm{cent}}^{\mathrm{LRG}}(M)$ to the satellite occupation function to statistically remove satellites from halos without centrals, effectively requiring a halo to have a central LRG before it can host satellite LRGs. \red{This is consistent with numerous HOD works such as \citet{2005Zheng, 2007bZheng} and more recently \citet{2015cGuo, 2020Alam}.}

\red{For ELGs, there has been several motivated HOD models mostly based off of semi-analytic models \citep[e.g.][]{2018Gonzalez-Perez, 2020Avila, 2020Alam, 2015Comparat}.} For this analysis, the baseline model refers to the skewed Gaussian model presented in \citet{2021bYuan}, which is based on the model presented in \citet{2020Alam}. We reproduce the baseline model here: 
\begin{align}
    N_{\mathrm{cent}}^{\mathrm{ELG}}(M)  &=  2 A \phi(M) \Phi(\gamma M)  + & \nonumber \\  
    \frac{1}{2Q} & \left[1+\mathrm{erf}\left(\frac{\log_{10}{M}-\log_{10}{M_{\mathrm{cut}}}}{0.01}\right) \right],  \label{eq:NHMQ}
\end{align}
where
\begin{align}
\phi(M) &=\mathcal{N}\left(\frac{\log_{10}M - \log_{10}{M_{\mathrm{cut}}}}{\sigma_M}\right), \label{eq:NHMQ-phi}\\
\Phi(\gamma M) &= \frac{1}{2} \left[ 1+\mathrm{erf} \left(\frac{\gamma(\log_{10}M - \log_{10}{M_{\mathrm{cut}})}}{\sqrt{2}\sigma_M} \right) \right], \label{eq:NHMQ-Phi}\\
A &=p_{\rm max}  -1/Q.
\label{eq:alam_hod_elg}
\end{align}
where $\mathcal{N}(x)$ represents a normalized unit Gaussian. The satellite occupation adopts a power law form,
\begin{align}
    N_{\mathrm{sat}}^{\mathrm{ELG}}(M) & = \left[\frac{M-\kappa M_{\mathrm{cut}}}{M_1}\right]^{\alpha}.
    \label{equ:elg_sat}
\end{align}
Note that compared to the LRG satellite HOD, we have removed the $N_\mathrm{cent}$ modulation term halo without central ELGs can still host satellite ELGs, contrary to the LRGs. This baseline HOD form for ELGs is also confirmed in simulation and semi-analytic model approaches by studies such as \citet{2021eHadzhiyska} and \citet{2020Gonzalez-Perez}. \red{For the mock ELGs, we do not introduce the $f_\mathrm{ic}$ incompleteness parameter. The ELG incompleteness is already implicitly encoded in the $p_\mathrm{max}$ and $M_1$ parameters, which respectively control the amplitude of the central and satellite occupation functions.}

Both of these baseline models assume that the halo occupation depends solely on halo mass. To attain the HOD of the mock galaxies, we first group each galaxy sample by the mass of the DMO host halo in logarithmic bins. Then for each halo mass bin, we tabulate the number of halos and the number of galaxies within that bin. The ratio of the two numbers gives the mean HOD of the sample as a function of halo mass. To differentiate between central and satellite galaxies, for each DMO host halo, we designate the galaxy associated with its most massive subhalo as the central, and the rest as satellites. 

\red{Figure~\ref{fig:lrg_hod} shows the comparison between the measured HOD of the LRG mocks in blue and orange dots and a fiducial baseline HOD model (Equation~\ref{equ:zheng_hod_cent}-\ref{equ:zheng_hod_sat}) with parameters tuned to match the measurement, as shown with the dashed lines. The parameter values of the fiducial model shown with the dashed lines are identified via a grid search, where we select the parametrization that best visually matches the measurement. We forego a full maximum likelihood analysis because the true HOD parameters are rather uncertain and dependent on the detailed physics prescriptions in the simulation, halo finding, target selection and more. The fiducial HOD values shown should thus be interpreted with generous error bars and only serve a rough reference for future studies.}

Nevertheless, it appears that the 5-parameter baseline HOD model can reproduce the mass dependence of the mocks reasonably well. Specifically, the centrals (shown in blue) do follow an error function shape at low masses before flattening out at $N = f_\mathrm{ic}$. The satellites (shown in orange) can be described reasonably well with a power law at higher masses and an extra decay term at lower masses. The drop-off in central occupation at higher halo masses is likely due to the limited sample size. 
For reference, the tuned parameters of the fiducial model are $\log M_\mathrm{cut} = 12.7$, $\log M_1 = 13.6$, $\sigma = 0.2$, $\alpha = 1.15$, $\kappa = 0.08$, and $f_\mathrm{ic} = 0.8$. In terms of derived HOD quantities, the selected LRG sample has an average halo mass per galaxy of $\bar{M}_h = 2.4\times 10^{13}h^{-1}M_\odot$, and a satellite fraction of $f_\mathrm{sat} = 20\%$. \red{Compared to CMASS LRG HODs that we derived in \citet{2021bYuan}, we find the DESI mock LRGs to have somewhat lower $M_\mathrm{cut}$ and $M_1$, corresponding to a lower typical halo mass and lower linear bias. This is to be expected given the higher redshift and higher number density of the expected DESI LRG sample.}

Figure~\ref{fig:elg_hod} showcases the comparison between the HOD of DESI-like mock ELGs in \textsc{IllustrisTNG} and a fiducial baseline HOD model with parameters tuned to match the measured HOD via a grid search. The baseline model refers to the skewed Gaussian model summarized in Equation~\ref{eq:alam_hod_elg} and Equation~\ref{equ:elg_sat}, \red{also known as the High Mass Quenched (HMQ) model.} The mock HOD is shown in dots, whereas the best-matching fiducial model is shown with the dashed line. The fiducial model parameters are chosen to match the mock measurements. Just as with the LRGs, the mass dependence in the ELG HOD can be well described by the baseline model. We note that the measured central occupation at $\log M > 13.5$ suffers from small number statistics as there are very few high mass halos hosting ELGs given the limited simulation volume. 
For reference, the model parameters of our fiducial model are $p_\mathrm{max} = 0.075$, $Q = 95$, $\log M_\mathrm{cut} = 11.9$, $\sigma = 0.5$, $\gamma = 5$, $\log M_1 = 14.2$, $\alpha = 0.65$, and $\kappa = 1.35$. In terms of derived HOD quantities, the ELG sample has a mean halo mass per galaxy of $6.0\times 10^{12}h^{-1}M_\odot$ and a satellite fraction of $33\%$. This is consistent with the expectation that ELGs are largely star-forming galaxies living in less massive halos. Compared to the mock LRGs, the higher satellite fraction of the mock ELGs indicates that ELGs living in more massive halos ($M_h > 10^{13}h^{-1}M_\odot$) are likely recently captured and have yet to quench or merge with the central. \red{Comparing to Table~1 of \citet{2020Alam}, which lists the best-fitting ELG HOD parameters derived from eBOSS clustering, we find good agreement in $M_\mathrm{cut}$, $\gamma$m, and $\sigma$, suggesting consistent central occupations between our mock ELG sample and the eBOSS best-fit. The satellite occupation is different, but that is at least partially due to differences in halo definition (\citealt{2020Alam} used \textsc{Rockstar} halos).}

\red{Both the mock LRG and ELG satellite fractions we found are approximately 30$\%$ larger than those found in previous BOSS and eBOSS studies. For BOSS LRGs, \citet{2021bYuan} found a marginalized satellite fraction of 11-15$\%$ depending on the HOD prescription. \citet{2017Zhai} found a similar LRG satellite fraction of $13\%$. For eBOSS ELGs, \cite{2019Guo} found a satellite fraction of 13-17$\%$, whereas \citet{2016Favole} found a satellite fraction of $22.5\pm2.5\%$. One potential reason for the higher satellite fraction in our study is the over-linking of the FoF halo finder, which is a tendency of FoF finder to connect neighboring halos that are otherwise physically distinct. This over-linking tendency would result in centrals in the vicinity of larger halos being identified as satellites. We discuss this effect in detail in Section~\ref{subsec:radial_bias}. Another effect that can contribute to higher satellite fractions in this study is differences in target selection between DESI and previous surveys. Finally, \citet{2020Avila} found with eBOSS data that the inferred ELG satellite fraction varies with the assumed HOD model. Specifically, they found the inferred satellite fraction to vary from $20-50\%$.}

\subsection{Satellite PDF}
\label{subsec:poisson}
A key assumption of the baseline HOD model is that the satellite occupation follows a Poisson distribution around the mean prediction. We test this assumption by tabulating the number of satellite LRGs and ELGs in our mock sample per halo mass bin. Then we compare the mean and variance of the number of satellites within each mass bin. If the distribution is indeed Poissonian, then we expect the mean and variance to be equal. 

\begin{figure}
    \centering
    \hspace*{-0.6cm}
    \includegraphics[width = 3.4in]{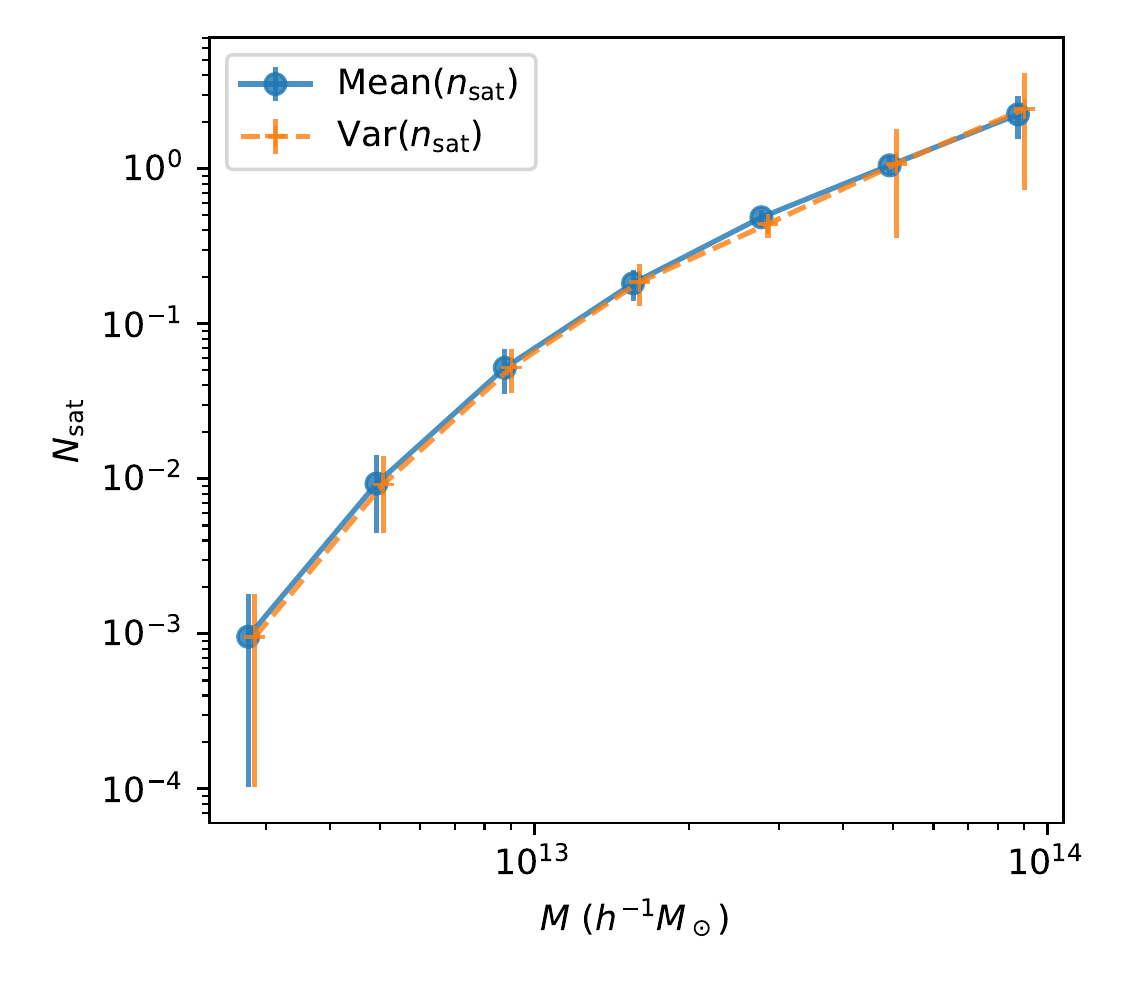}
    \vspace{-0.3cm}
    \caption{\red{The mean and variance of the mock LRG satellite occupation as a function of halo mass to check for potential deviations from a Poisson sampling of satellites.} The errorbars come from jackknife resampling of the simulation volume. We see that the mean and variance are equal to remarkable accuracy, consistent with the hypothesis that the satellite occupation follows a Poisson distribution. }
    \label{fig:lrg_poisson}
\end{figure}

\begin{figure} 
    \centering
    \hspace*{-0.6cm}
    \includegraphics[width = 3.4in]{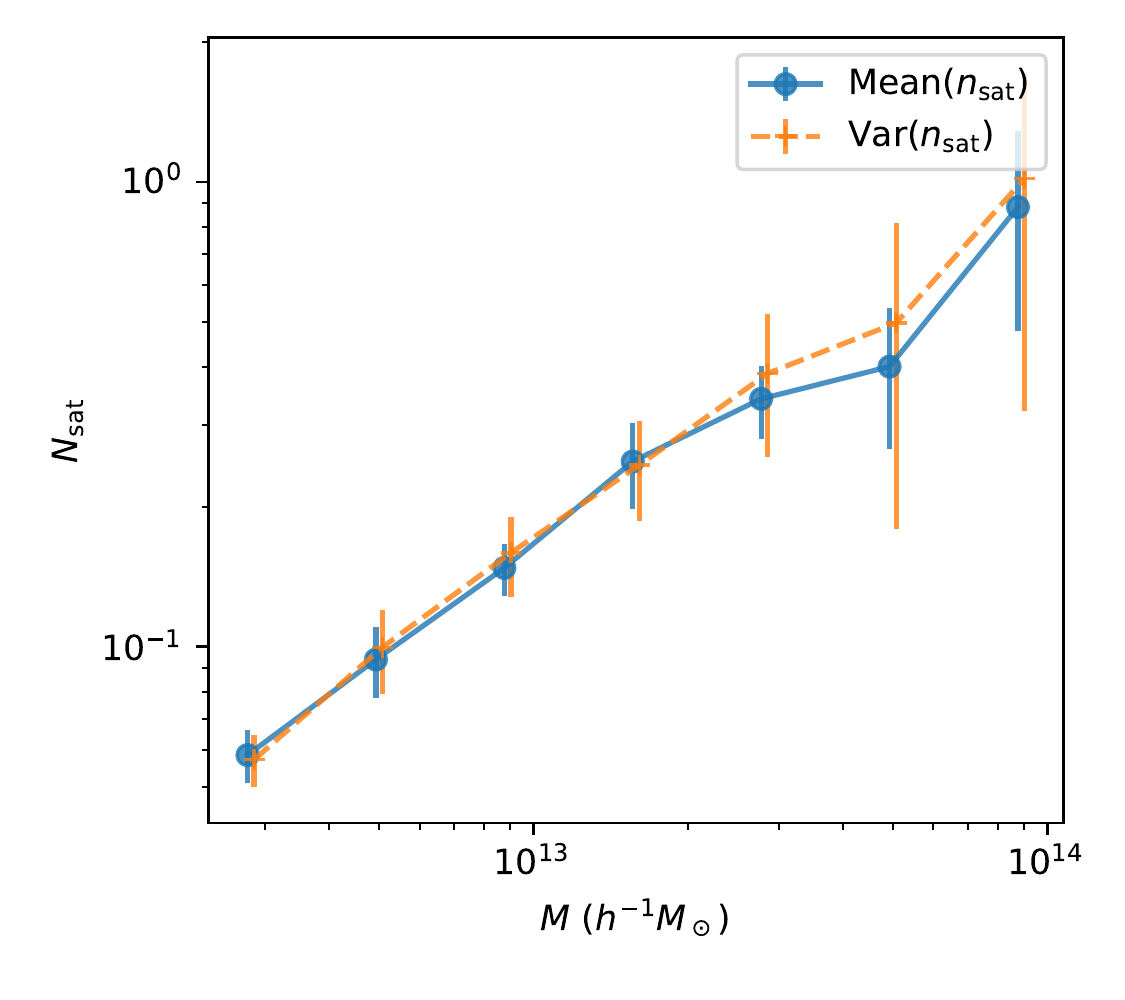}
    \vspace{-0.3cm}
    \caption{\red{The mean and variance of the mock ELG satellite occupation as a function of halo mass to check for potential deviations from a Poisson sampling of satellites.} The errorbars come from jackknife resampling of the simulation volume. We see that the mean and variance are equal within uncertainty, consistent with the hypothesis that the satellite occupation follows a Poisson distribution. }
    \label{fig:elg_poisson}
\end{figure}

Figure~\ref{fig:lrg_poisson} showcases the mean and variance of mock LRG satellite occupation, as a function of halo mass. The errorbars come from jackknife resampling of the simulation volume. We see that the mean and variance perfectly agree with each other across the full halo mass range, consistent with the assumption that LRG satellite occupation follows a Poisson distribution. 
Figure~\ref{fig:elg_poisson} showcases the mean and variance of mock ELG satellite occupation, as a function of halo mass. Again the errorbars are computed via jackkife resampling of the simulation volume. We see that the mean and the variance are consistent across most of the mass range. At the largest halo mass, the variance appears to supersede the mean, potentially pointing to the ELG satellites having a super-Poisson distribution. However, this difference is not statistically significant compared to the amount of uncertainty. 

Nevertheless, the fact that the ELG satellite occupations are potentially super-Poisson has interesting physical implications. From a galaxy formation perspective, if satellites formed and evolved in sub-halos independently, then we would expect the satellite occupation to be Poissonian. However, if the satellites within the same halo have correlated formation history, also known as 1-halo conformity, then their occupation would tend towards being super-Poisson. Satellite conformity can also arise if satellite evolution depends on external properties such as the local environment, which is one of the main effects of interest in this analysis. In fact, we show in Section~\ref{subsec:fenv} that ELG satellite occupation indeed correlates with the local environment. 

\red{To put this result in the broader context, previous studies have found in simulations that satellite hosting subhaloes follow a distribution that is either close to Poisson or super-Poisson \citep[e.g.][]{2017Jiang, 2009Boylan-Kolchin}. More recently, \citet{2019Jimenez} found that semi-analytical star-forming satellite galaxies are best described by a super-Poisson negative binomial distribution. However, \citet{2020Avila} found that the satellite PDF is in fact degenerate with other HOD modeling choices, and that clustering alone is not sufficient in breaking such degeneracy.} 

Finally, we note the caveat that the our moment test is a necessary but not sufficient condition for a Poisson distribution. In principle, we can extend the presented analysis to higher moments for additional constraining power. In fact, we have calculated the third moment of the satellite PDF, but the error bars on the third moments are too large to be informative given the limited sample size. We reserve a more thorough analysis on the satellite occupation PDF, potentially through a Kolmogorov-Smirnov test or an Anderson-Darling test, for a future study when a larger mock sample becomes available. 

\subsection{Radial biases}
\label{subsec:radial_bias}
While the mass dependence of the mock LRG and ELG HOD seems to be well described by the baseline HOD models, in this and the following subsection, we explore whether other assumptions of the baseline HOD model are also upheld. 

One common assumption in HOD analyses is that the distribution of satellite galaxies within the dark matter halo follows the distribution of dark matter itself, or in many cases an NFW profile. We test this assumption by splitting the mock galaxies into halo mass bins, and within each mass bin, comparing the stacked radial distribution of satellites to the halo mass profile. The average halo mass profile is obtained by conducting NFW fits to the DMO mass profile of each halo in the mass bin, and then averaging over the best-fits. 

\begin{figure*}
    \centering
    \hspace*{-0.6cm}
    \includegraphics[width = 6.3in]{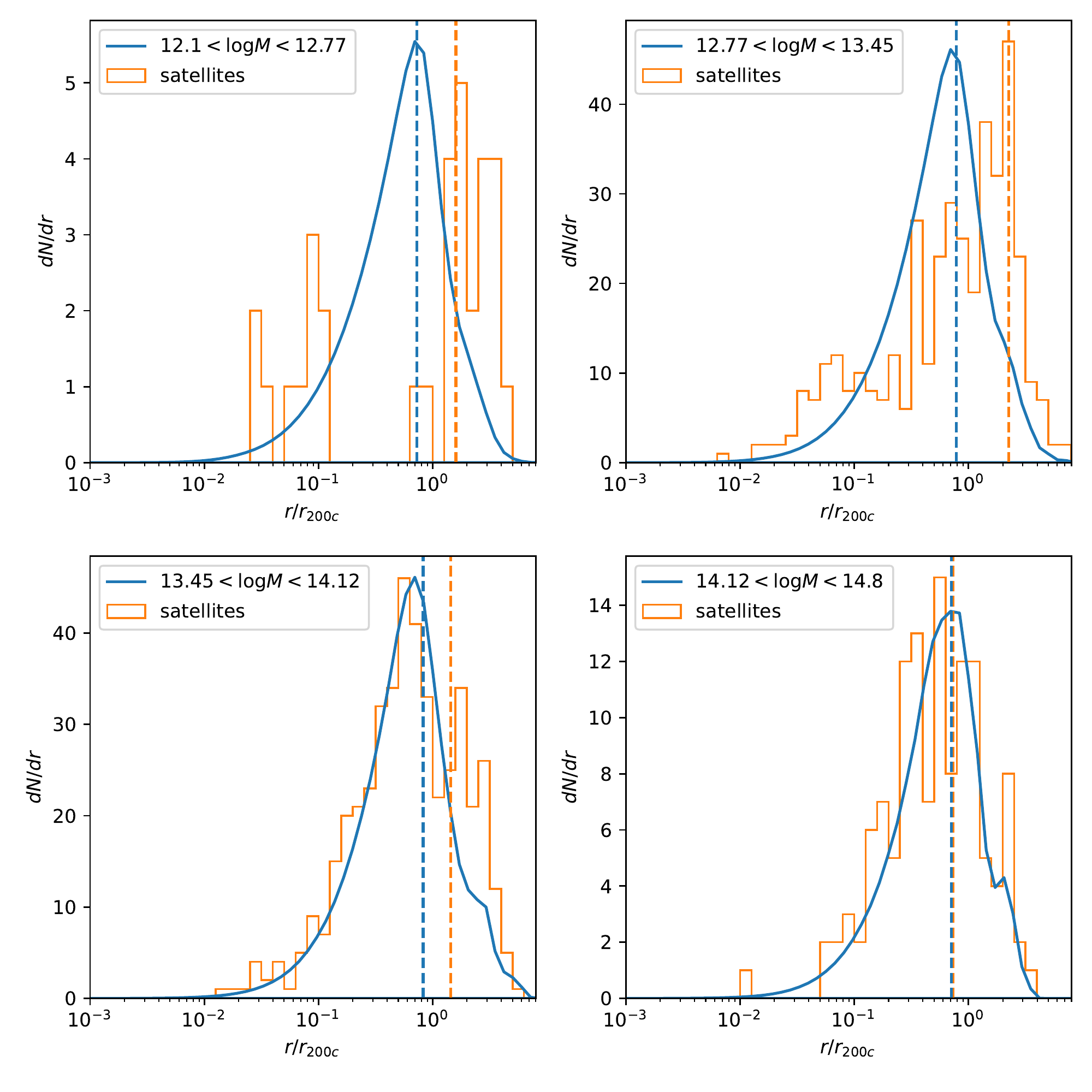}
    \vspace{-0.3cm}
    \caption{\red{The radial distribution of mock LRG satellites compared to the halo total mass profile. The mock LRG satellites are split to 4 mass bins. The halo mass profile, shown in blue, is compiled by stacking the NFW profile best-fit of each halo within the mass bin. The $x$-axis shows the radial position relative to the center of the central subhalo, normalized by $r_{200c}$ of the halo. The $y$-axis shows the number of satellites per radial bin. We have normalized the mass profile by an arbitrary factor for easy visualization. The vertical dashed lines show the mean of the radial distributions. We see clear differences in the radial distribution in the mock LRG satellites from the halo mass profile. We speculate that the under-abundance of satellites at larger radii is at least partially due to the over-linking in the FOF halo finder. See Section~\ref{subsec:radial_bias} for a detailed discussion.} }
    \label{fig:lrg_radial}
\end{figure*}

Figure~\ref{fig:lrg_radial} showcases the radial bias of the mock LRGs, where we show the satellite radial distribution in orange and halo mass profiles in blue. \red{The mean of the distributions are plotted with the vertical dashed lines. The matter profiles shown in blue have been normalized to approximately the same height as the satellite distribution for ease of visualization.} We see a clear mass-dependent deviation in the satellite radial distribution relative to the dark matter halo. Specifically, the satellite galaxies appear to preferentially occupy the outskirts of halos, resulting in a second peak in satellite radial distribution beyond the halo radius $r_{200c}$. This radial bias dissipates at the highest halo mass, largely disappearing at $M>10^{14.1}h^{-1}M_\odot$. We speculate that many of the satellites far from halo centers are ``mis-assigned'', in that they are in fact centrals of their own halos or physically associated with neighboring more massive halos. This is an expected issue given the FOF algorithm's tendency to over-link halos, where either physically distinct halos are merged into one, or a halo in the vicinity of other halos claims a subset of neighbors' particles. One immediate piece of evidence for this explanation is that, for lower mass halos below $\log M < 13$, most of these halos are not expected to host satellites (refer to Figure~\ref{fig:lrg_hod}). Thus, these satellites are in fact centrals of nearby less massive halos. 

\begin{figure}
    \centering
    \hspace*{-0.6cm}
    \includegraphics[width = 3.5in]{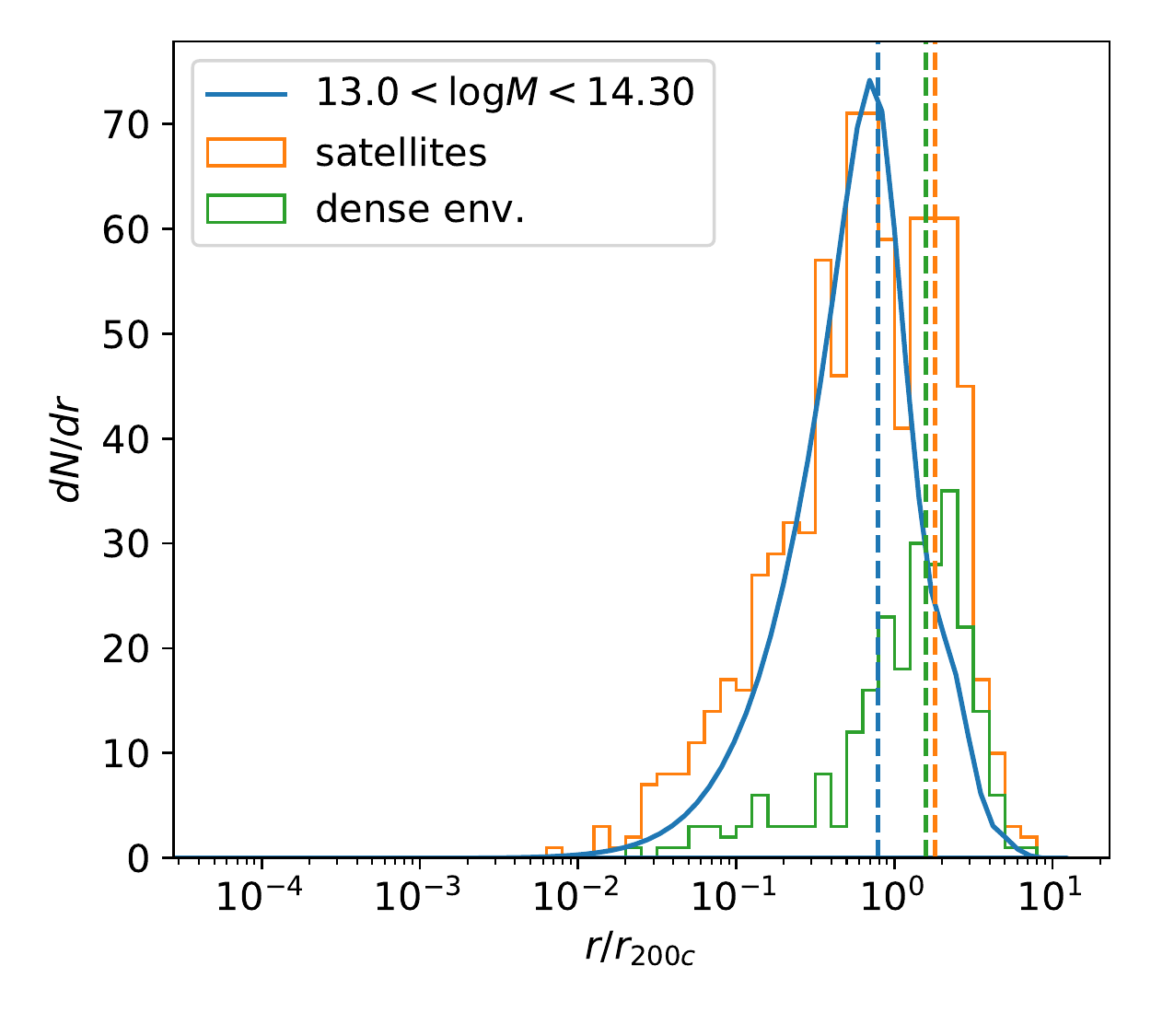}
    \vspace{-0.3cm}
    \caption{The radial distribution of the mock LRG satellites compared to the halo mass profile, in the $13<\log_{10}M<14.3$ mass range. The additional green histogram showcases the radial distribution of satellites above a local environment threshold, which is chosen to select the top $30\%$ satellites in terms of environment \red{(conceptually the total mass of neighboring halos within $5h^{-1}$Mpc, defined more rigorously in Equ.~\ref{equ:fenv})}. The dashed vertical lines represent the mean of the three distributions. The satellites in denser environments appears to occupy outskirts of host halos beyond $r_{200c}$, consistent with an over-linked halo explanation. }
    \label{fig:overlink}
\end{figure}

Another way we test this explanation is by examining LRG satellites in denser environments, where we expect halo over-linking to occur most frequently. As a result, we expect satellites in denser environments to more likely occupy over-linked halos, thus showing a radial distribution that is skewed further away from the halo center. Figure~\ref{fig:overlink} illustrates this test, where we show the same kind of radial distribution comparison as Figure~\ref{fig:lrg_radial}, albeit in a new custom mass bin, but we also over-plot the radial distribution of satellites above a certain environment threshold (selects top $30\%$ in environment) in green. The environment is defined as the enclosed mass of neighboring halos within a $5h^{-1}$Mpc radius, roughly tracking the local density in the large scale structure. The figure confirms that the satellites in denser environments preferentially occupy outer regions of their host halos. Since their radial positions are beyond $r_{200c}$, these satellites most likely occupy over-linked halos. In fact, the distribution of the green subsample perfectly coincides with the excess in the full satellite sample (orange) relative to the dark matter distribution. It is also interesting that the green subsample does not show a significant number of galaxies below $r_{200c}$, despite the fact that we expect halos in this mass range to host on the order of 1 to a few satellites. This suggests that the underlying physically bound halos are in fact lower in mass, but up-scattered into the designated mass range by over-linking with additional objects. All in all, this is consistent with our explanation that the over-abundance of satellites at large radii is due to preferential occupation of over-linked halos in dense environments. \red{To confirm this explanation, we propose a re-analysis of these samples in a future paper with different halo finders, such as \textsc{Rockstar} \citep{2013Behroozi}, which avoids over-linking by utilizing velocity information in additional to positional information.}

Nevertheless, these trends clearly break from the common assumptions of HOD implementations, where satellites are assumed to follow the dark matter distribution, regardless of halo finding algorithm and halo mass. This supports the need for flexibilities in modeling the satellite distribution and its mass dependency. While this radial bias likely only affects projected clustering on small scales, it could significantly bias the velocity distribution of satellite galaxies and thus has significant impact on the predicted redshift-space distortion signature out to $\sim 50h^{-1}$Mpc along the LOS. 

\begin{figure*}
    \centering
    \hspace*{-0.6cm}
    \includegraphics[width = 6.3in]{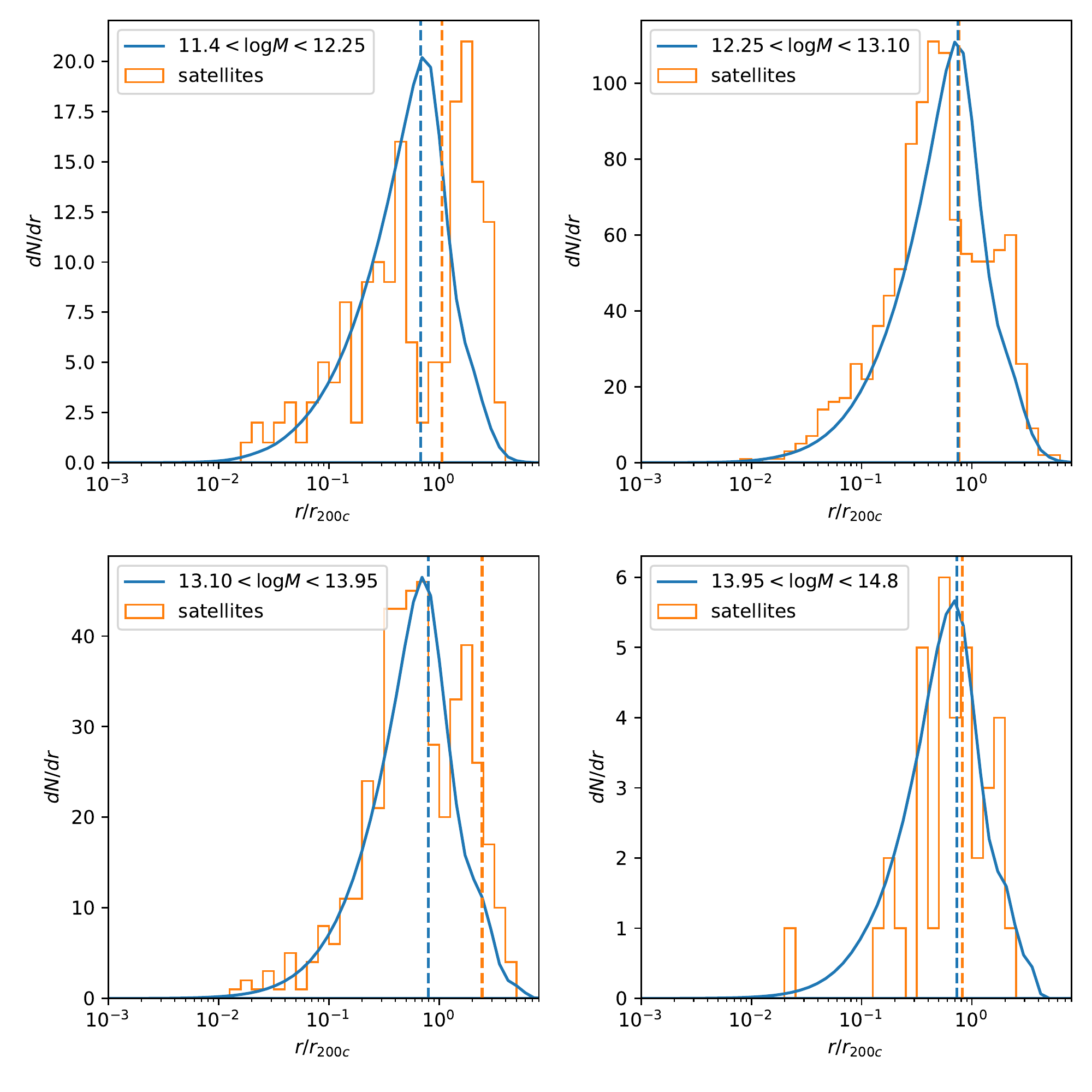}
    \vspace{-0.3cm}
    \caption{\red{The radial distribution of mock ELG satellites compared to the halo mass profile. Similar to the setup of Fig.~\ref{fig:lrg_radial}, the mock ELG satellites are split to 4 mass bins. The mass profile, shown in blue, is compiled by stacking the NFW profile best-fit of each halo within the mass bin. The vertical dashed lines show the mean of the radial distributions. Similar to the mock LRGs, there is a clear bimodal behavior at lower halo masses. This can be again explained by the over-linking in the FOF halo finders.}}
    \label{fig:elg_radial}
\end{figure*}

Figure~\ref{fig:elg_radial} showcases the radial bias of the mock ELGs, where we show the satellite distribution in orange and the normalized halo mass profiles in blue. Again the halo mass profile comes from averaging the NFW fits of DMO halos in the mass bin. We see indications of a bimodal trend in satellite distribution that perhaps suggests the existence of two distinct populations of ELG satellites. One population preferentially occupy the outskirts of low mass halos, whereas the other population largely follows the mass profile of the halos, albeit slightly more concentrated. While we speculate that the outer population is again due to over-linked halos, we should mention that several previous findings did find ELGs to potentially prefer the outskirts of halos \citep[e.g.][]{2016Alpaslan, 2018Orsi, 2020Avila}. 

Another interesting scenario that could contribute to this outer population in both LRGs and ELGs is related to the concept of splashback radius, which is a more physically-motivated halo boundary definition that is often 2-3 times larger than the canonical virial radius definition \citep[e.g.][]{2014Diemer, 2015bMore, 2016More}.
Specifically, we find that the outer population of ELGs tend to occupy halos in denser environment compared to the inner population, and that the outer population of ELGs tend to have lower specific star formation rate compared to the inner population. This is potentially consistent with the notion that these outer galaxies are possibly splashback galaxies that had some of their gas stripped during encounters with nearby massive halos. All in all, while we speculate that imperfections in the halo finder to account for much of this phenomenon, the radial excess of satellites could also be reflective of other physical processes. 

The inner peak at lower halo mass for ELGs is also interesting in that it could be connected to boosted star formation rate among near neighbors. Specifically, many studies have found evidence of increased star formation rate among late-type galaxies in dense environments \citep[e.g.][]{2011Wong, 2011Patton, 2013Patton, 2019Moon}, often explained by a combination of the tidal effect of neighboring galaxies and other hydrodynamic effects. We refer curious readers to the above citations for descriptions of recent findings.

\subsection{Velocity biases}
In this subsection, we investigate the velocity biases of the DESI-like LRG and ELG samples. Velocity biases generally refer to the phenomenon where the velocity of the central and satellite galaxies differ from that of the underlying dark matter. This effect, while not relevant for projected clustering, is essential in modeling the small-scale clustering in redshift-space, as showcased in several recent full-shape clustering analyses of BOSS galaxies, such as \citet{2015aGuo, 2021bYuan}. Within the HOD framework, velocity bias manifests as a central velocity deviation from the central subhalo and a satellite velocity deviation from its host particle.  Mathematically, this is quantified by the central and satellite velocity bias parameters $\alpha_c$ and $\alpha_s$,
\begin{align}
    \alpha_c & = \frac{\sigma_{\mathrm{pec, cent}}}{\sigma_\mathrm{halo}}, \\
    \alpha_s & = \frac{\sigma_{\mathrm{pec, sate}}}{\sigma_\mathrm{halo}},
    \label{equ:velbias}
\end{align}
where $\sigma_{\mathrm{pec, cent}}$ is the central peculiar velocity dispersion, $\sigma_\mathrm{halo}$ is the halo velocity dispersion, and $\sigma_{\mathrm{pec, sate}}$ is the satellite peculiar velocity dispersion. The galaxy peculiar velocity is defined as the galaxy velocity minus the halo velocity, which is computed as the weighted average of the halo's particle velocities. By this definition, if there is no velocity bias, then the central peculiar velocity would be zero, whereas the satellite velocity would track that of the dark matter particles. Thus, no velocity bias corresponds to $\alpha_c = 0$ and $\alpha_s = 1$. A series of studies have found that in order to reproduce the observed BOSS LRG clustering, values of $\alpha_c \approx 0.2$ and $\alpha_s \approx 1$ are preferred \citep[e.g.][]{2021Yuan, 2015aGuo}. This corresponds to a model where the peculiar velocity of centrals relative to central subhalo is $20\%$ of the halo velocity dispersion, and satellite peculiar velocity dispersions relative to halo center is slightly less than its host particle. Most recently, a thorough analysis of the CMASS redshift-space distortion signal finds $\alpha_c = 0.18^{+0.03}_{-0.04}$ and $\alpha_s = 1.00^{+0.03}_{-0.03}$ for the CMASS LRG sample \citep[][]{2021bYuan}. 

We can test for velocity bias in our LRG and ELG mocks by directly comparing the galaxy velocities to the velocity dispersions of the host halo and subhalos. The only technical difficulty in this calculation is that the velocity dispersions of the halos are not readily available in the \textsc{IllustrisTNG} data products, and pulling the particle subsample for each host halo is a relatively expensive operation. Thus, we approximate the halo velocity dispersion by the velocity dispersion of its largest subhalo. We test the bias due to this approximation in a randomly selected set of 800 halos across the full halo mass range. For each halo, we extract its particle sample from the full simulation and compare their velocity dispersion with that of the largest subhalo. We find that the largest subhalo velocity dispersion systematically over-estimates the halo velocity dispersion by $(23\pm 3)\%$, without any significant mass dependence. This makes sense as the largest subhalo consists of particles deeper in the halo potential well. Thus, we approximate the halo velocity dispersion as the velocity dispersion of the largest subhalo divided by 1.23. Finally, we measure velocity bias by taking the ratio between central/satellite peculiar velocities and the halo velocity dispersion. 

\begin{figure}
    \centering
    \begin{subfigure}[]{0.4\textwidth}
    \hspace*{-0.8cm}
    \includegraphics[width = 3.3in]{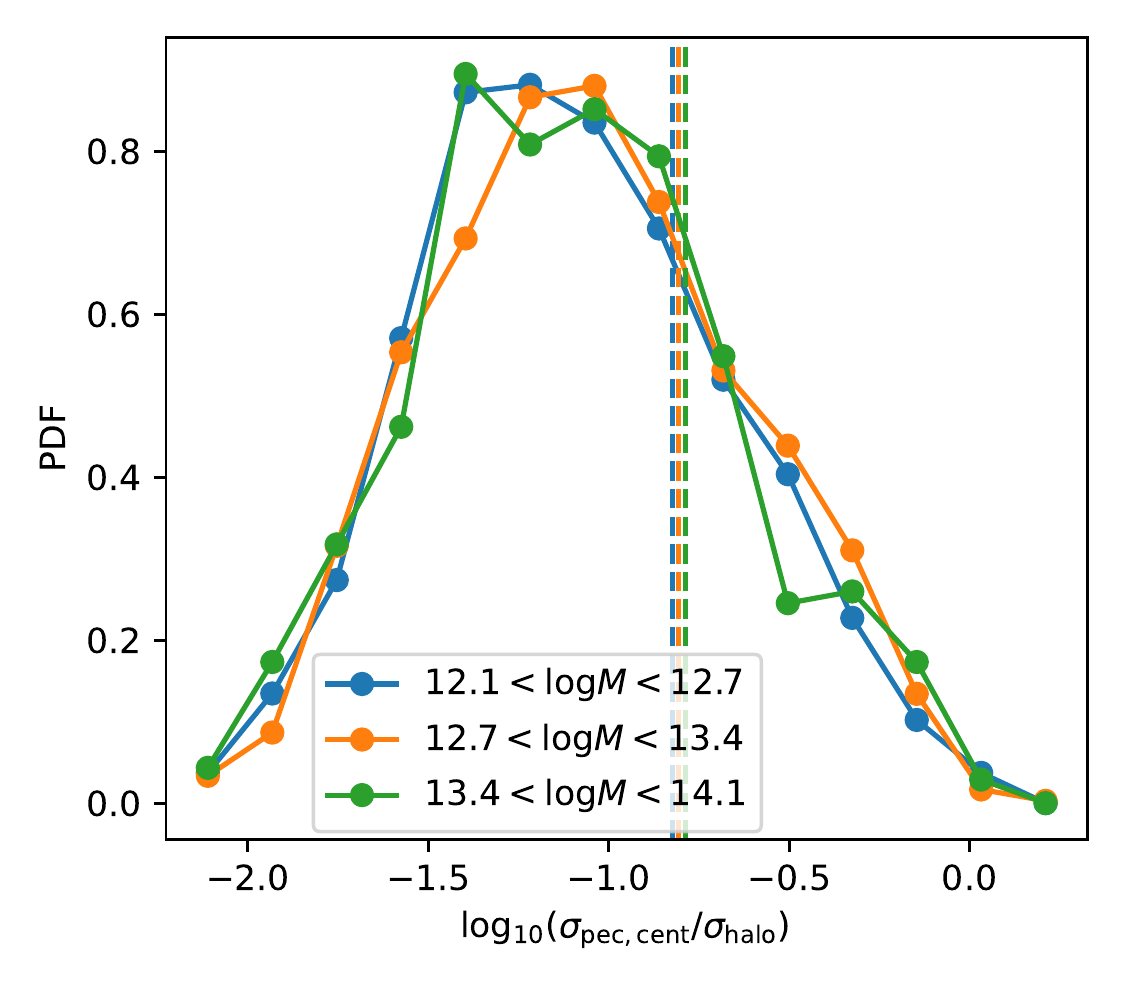}
     \vspace{-0.5cm}
    \caption{LRG central velocity bias}
    \label{fig:lrg_vel_cent}
    \end{subfigure}
    \begin{subfigure}[]{0.4\textwidth}
    \hspace*{-0.8cm}
    \includegraphics[width = 3.3in]{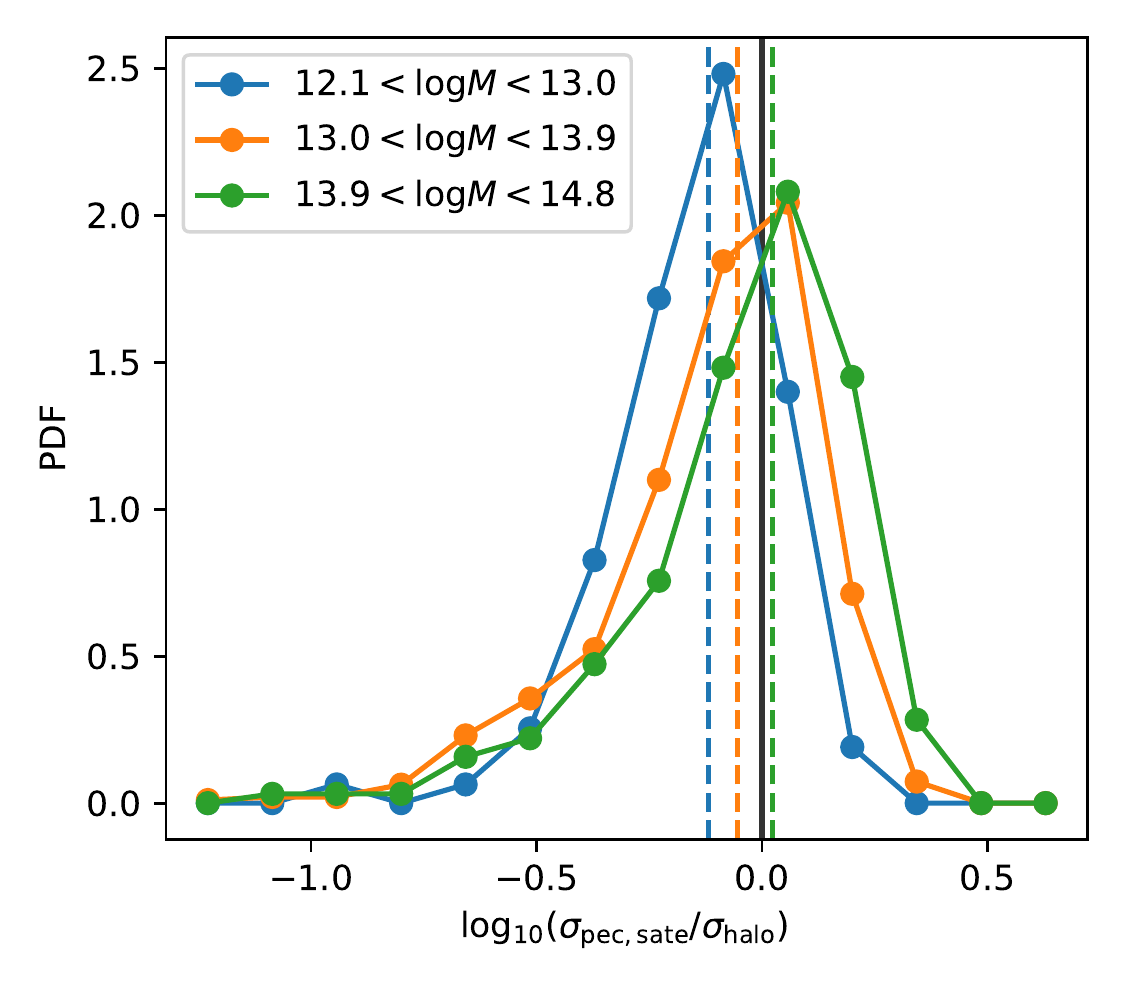}
     \vspace{-0.5cm}
    \caption{LRG satellite velocity bias}
    \label{fig:lrg_vel_sate}
    \end{subfigure}
    \caption{The velocity bias signature of the DESI-like LRG mock. The top and bottom panel corresponds to the centrals and satellites, respectively. Each panel shows the distribution of the ratio between galaxy peculiar velocities and the host halo velocity dispersion. The vertical dashed lines denote the corresponding central and satellite velocity bias parameters $\alpha_c$ and $\alpha_s$ for that mass bin. If no velocity bias, one would expect the ratio for the centrals to be 0 (negative infinity on the log scale), and the ratio for the satellites to be 1 (marked by the solid black line). We divide the centrals and satellites each into three mass bins to test for mass dependencies. For the centrals, we find the central velocity ratio to peak around $10\%$, with a central velocity bias parameter of around $0.14$ and no significant evidence for mass dependency. For the satellites, we find the velocity ratio to peak close to 1, with a satellite velocity bias parameter of $\alpha_s = 0.8, 0.9, 1.05$ for the three mass bins. The satellite velocity bias does seem to slightly increase with halo mass.}
    \label{fig:lrg_vel}
\end{figure}

Figure~\ref{fig:lrg_vel} showcases the velocity bias signature for the LRG mock, where we show the distribution of the ratio of the galaxy peculiar velocity to the halo velocity dispersion, with centrals on the top panel and satellites on the bottom panel. We divide each sample into three mass bins to test for mass dependency. The dashed vertical lines denote the corresponding velocity bias parameter for each mass bin. The centrals do display clear velocity bias signature, with a peak peculiar velocity to halo dispersion ratio of around 0.1, and a corresponding central velocity bias parameter $\alpha_c = 0.137\pm 0.007$. The errorbar comes from a jackknife division of the simulation volume. This is slightly smaller than the $\alpha_c = 0.2$ signature found for BOSS LRGs, but still statistically consistent to about 1$\sigma$. We do not find significant evidence for mass dependency in the central velocity bias. For the satellites, we find the peculiar velocity dispersion ratio to peak around 1, with a mean of 0.92. The inferred satellite velocity bias parameter is $\alpha_s = 0.92\pm 0.05$. The satellite velocity bias is slightly less than 1, but again in $1\sigma$ agreement with the results for BOSS LRGs. We do find evidence for mass dependency, as the mean velocity ratios in the three mass bins monotonically increase with halo mass. Specifically, we get 0.8, 0.9, and 1.05 in the three mass bins. 

\begin{figure}
    \centering
    \begin{subfigure}[]{0.4\textwidth}
    \hspace*{-0.8cm}
    \includegraphics[width = 3.3in]{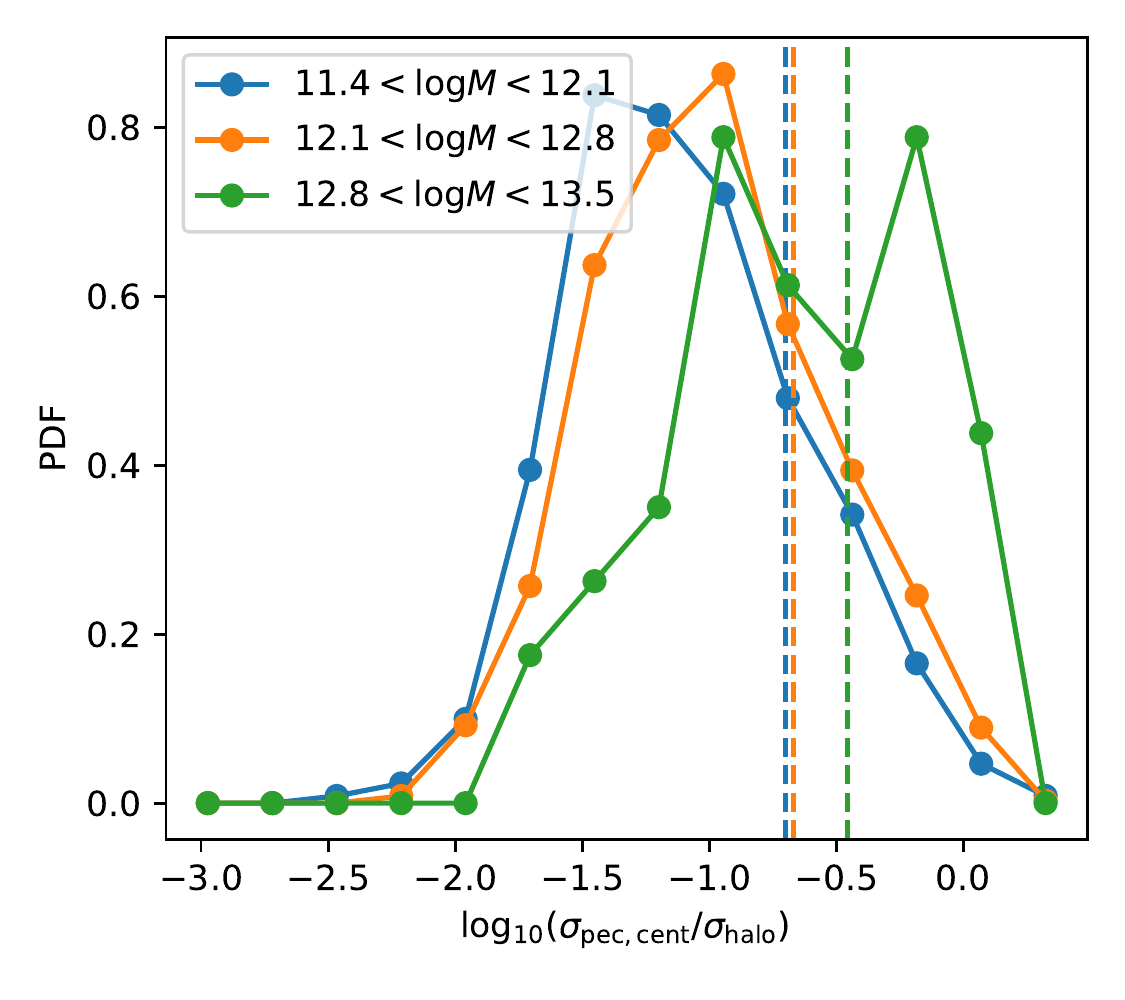}
     \vspace{-0.5cm}
    \caption{ELG central velocity bias}
    \label{fig:elg_vel_cent}
    \end{subfigure}
    \begin{subfigure}[]{0.4\textwidth}
    \hspace*{-0.8cm}
    \includegraphics[width = 3.3in]{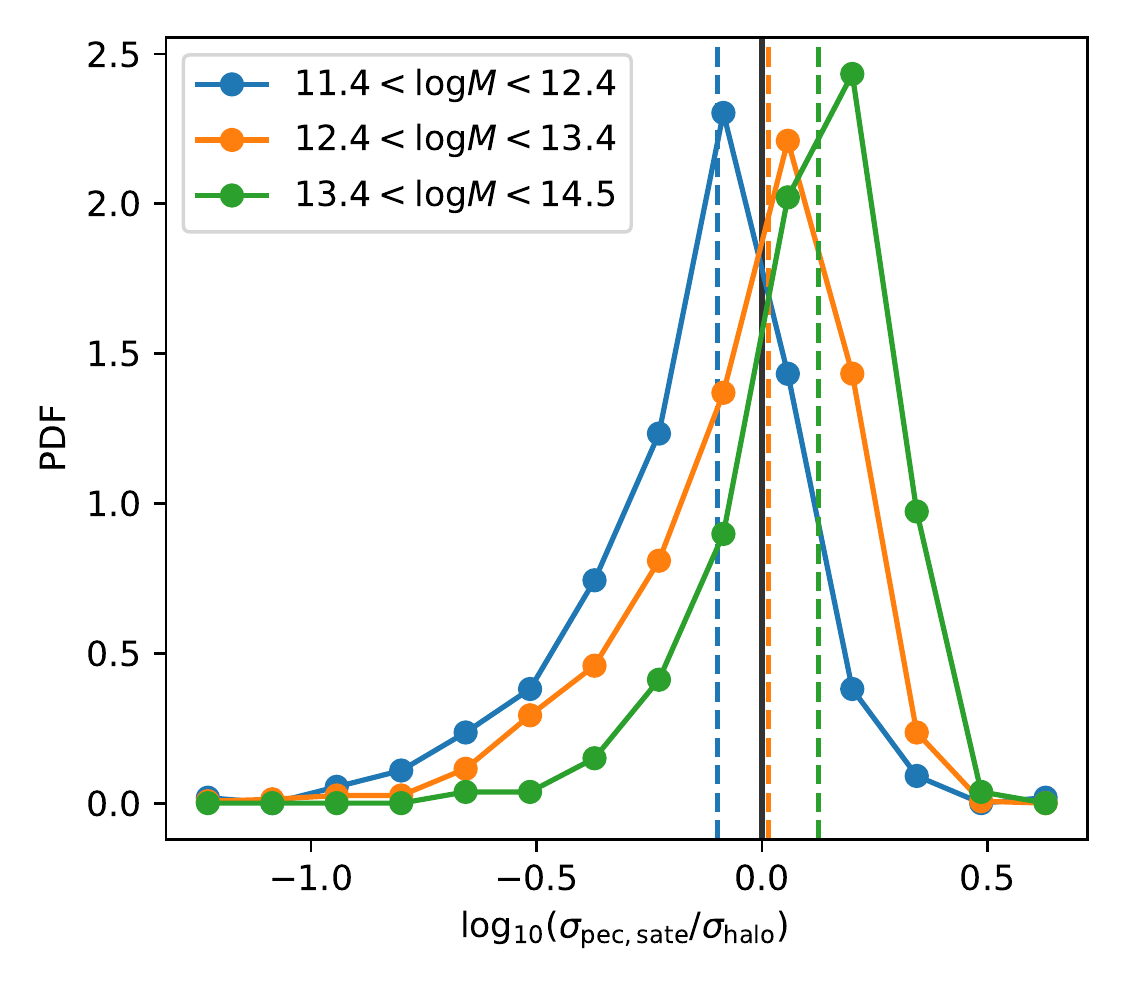}
     \vspace{-0.5cm}
    \caption{ELG satellite velocity bias}
    \label{fig:elg_vel_sate}
    \end{subfigure}
    \caption{The velocity bias signature of the DESI-like ELG mock. The top and bottom panel corresponds to the centrals and satellites, respectively. Each panel shows the distribution of the ratio between galaxy peculiar velocities and the host halo velocity dispersion. The dashed lines denote the corresponding velocity bias parameter. We divide the centrals and satellites each into three mass bins to test for mass dependencies. We find the central velocity bias to peak around $10\%$ and a potential dependency on halo mass. The bimodal distribution of central peculiar velocity in mass bin $12.8 < \log M <13.5$ is likely due to noise. The mean central velocity bias parameter is around 0.15. The satellite peculiar velocity ratios is not significantly different from 1, but the distributions show a clear mass dependency. The satellite velocity bias parameters for the three mass bins are $\alpha_s = 0.88, 1.02$, and 1.13, respectively. }
    \label{fig:elg_vel}
\end{figure}

Figure~\ref{fig:elg_vel} showcases the velocity bias signature for the ELG mock, where again we plot separately the centrals and satellites on the top and bottom panels, respectively. Similar to the mock LRGs, we see clear velocity bias signature for the centrals, where the peculiar velocity distributions peak around 0.1, and the central velocity bias parameter is $\alpha_c = 0.163\pm 0.010$. The satellite velocity dispersion ratios do not show significant deviation from 1, with an inferred satellite velocity bias parameter of $\alpha_s = 1.01\pm 0.04$. However, the mock ELG satellite peculiar velocity ratios exhibit clear mass dependencies. Specifically, mock ELGs in more massive halos tend to show higher velocity biases, with $\alpha_s = 0.88, 1.02$, and 1.13 in the three mass bins, respectively. This can potentially be explained by the fact that ELGs in massive halos tend to be young and associated with recent mergers, which means that they tend to be less virialized and exhibit large peculiar velocities. \red{This compares to \citet{2018Orsi}, where two distinct populations of ELG satellites were idenitified using a semi-analytic catalog, one recently accreted populating the outskirts of the host halos and showing a large infall velocity, and the other undergoing gas-stripping processes but still hosting sufficient star formation to be included in ELG selections. The mock ELG satellites we find in higher mass halos are consistent with the first population as they both display larger than unity velocity biases. \citet{2020Avila} introduced a satellite velocity bias parameter in their analysis of eBOSS clustering, and found it to be degenerate with other HOD choices. They find that a value of $\alpha_s = 1$ or $\alpha_s > 1$ are often preferred.} 

For both the LRG and ELG mocks, we find evidence in support of the velocity bias model among the centrals. For the satellites, we find evidence for mass dependence in the velocity bias signal for both samples. Encouragingly our LRG results are consistent with that of BOSS LRG analysis. We find potential mass dependencies in satellite velocity bias for both the mock LRGs and ELGs. Our findings show that velocity bias is a critical ingredient in correctly modeling the full-shape galaxy clustering in BOSS and DESI. Another caveat to consider is the effect of the halo finding technique adopted in identifying halos and subhalos, subsequently affecting the inferred relative velocities of galaxies with respect to their perceived halo parent.

\subsection{Secondary occupation biases}

In this subsection, we examine the validity of the mass-only    assumption of the baseline HOD models when applied to the DESI-like mock galaxies. Specifically, the baseline HOD models assume that galaxy occupation in a dark matter halo only depends on halo mass. Such an assumption has recently been challenged extensively in both observations and simulations, where we have been finding growing evidence for secondary dependencies on halo properties such as concentration and environment \citep[e.g.][]{2020Hadzhiyska, 2020Xu, 2021bHadzhiyska, 2021bYuan}. We generically refer to these secondary dependency as secondary galaxy bias. The term galaxy assembly bias is sometimes used interchangeably, but it technically only refers to secondary properties related to the assembly history. With \textsc{IllustrisTNG}, we have direct access to the galaxy occupation and the associated halo properties, presenting an excellent opportunity to examine galaxy secondary bias in a realistic scenario. Before we proceed, however, we point out the important caveat that the detection of secondary galaxy biases is often dependent on halo definition. The fact that \textsc{IllustrisTNG} halos are defined differently than those we studied in previous works \citep[e.g.][]{2021Yuan, 2021bYuan} means that we could see differences in the measured secondary bias. 

\subsubsection{Concentration-based secondary bias}

Halo concentration has long been used as the standard tracer of galaxy secondary bias because it is easy to compute in a simulation and found to correlate strongly with halo assembly history \citep[e.g.][]{2002Wechsler, 2007Croton, 2007Gao}, with older halos having higher concentrations. In brief, the key intuition is that early forming halos have more time to undergo mergers and tidal disruption, thus resulting in fewer but more luminous galaxies. The majority of subsequent HOD-based analyses have utilized the halo concentration as the sole marker for galaxy secondary bias (assembly bias) \citep[e.g.][]{2014Zentner, 2016Hearin, 2019Lange}. 

In this subsection, we examine the strength of concentration-based secondary bias in our DESI LRG and ELG mock. We adopt the standard definition of halo concentration, 
\begin{equation}
    c = \frac{r_{200c}}{r_s},
\end{equation}
where $r_{200c}$ is the spherical overdensity radius of the halo, and $r_s$ is the scale radius that comes out of the NFW fit. We additionally normalize the concentration into a concentration rank, where we divide the halos into narrow mass bins (70 logarithmic bins between $1.6\times 10^{11}h^{-1}M_{\odot}$ and $6.3\times 10^{14}h^{-1}M_{\odot}$), and within each mass bin, we rank-order the halos by their concentration. We then normalize the ranks to within range -0.5 and 0.5 for ease of comparison across difference mass bins. The idea is look for evidence that the mean concentration rank of the galaxies might be greater or less than 0, because in the no-secondary bias case, the galaxies would equally prefer more concentrated and less concentrated halos. 

\begin{figure*}
    \centering
    \hspace*{-0.3cm}
    \includegraphics[width = 7.1in]{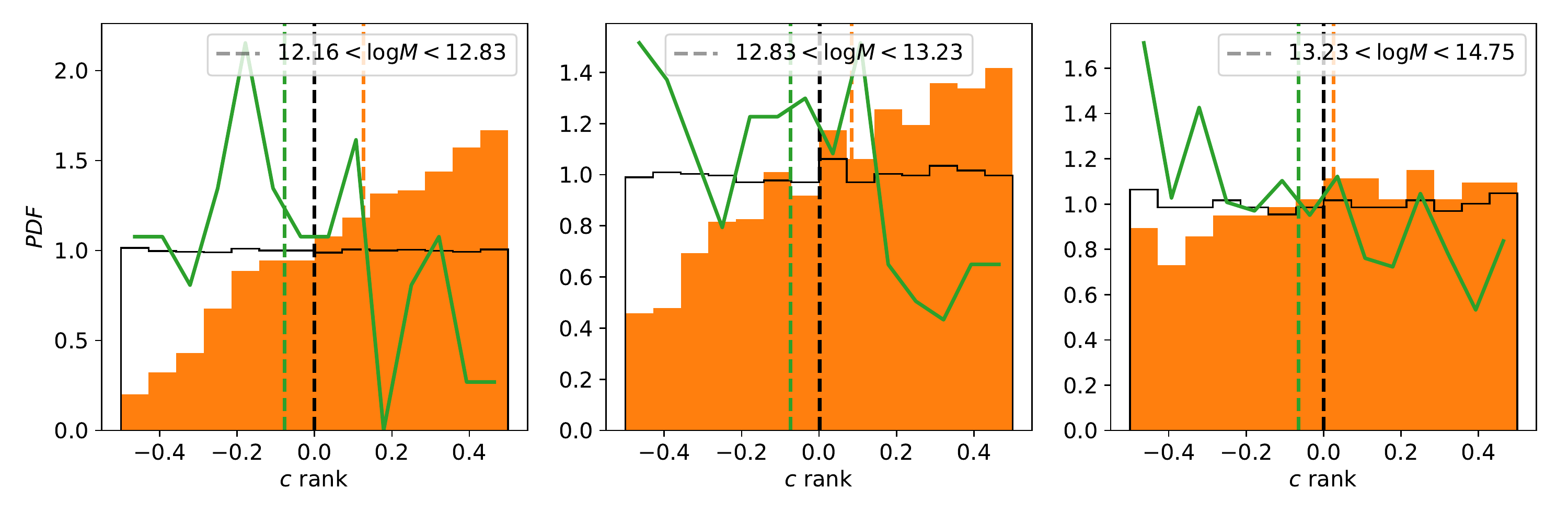}
    \vspace{-0.3cm}
    \caption{The halo occupation of DESI LRG mock galaxies as a function of concentration per mass bin. The $x$-axis shows the rank of halo concentration within a small halo mass bin, normalized to within $-0.5$ and 0.5. By definition, in each mass bin, $50\%$ of halos will have a positive $c$ rank whereas the other $50\%$ will have a negative $c$ rank. The black histogram shows the distribution of halos within the mass bin, showing a largely symmetric distribution around 0. The orange histogram shows the distribution of the halo concentration of the central galaxies, whereas the green curve shows the distribution of the satellite galaxies. All distributions are normalized to 1, so the relative magnitude of the distributions are not meaningful. The dashed lines represent the mean of the distributions. We see strong concentration-based dependencies, with centrals preferring more concentrated halos and satellites preferring less concentrated halos.}
    \label{fig:lrg_crank}
\end{figure*}

Figure~\ref{fig:lrg_crank} showcases the galaxy secondary bias signature of DESI LRG mock galaxies. Specifically, it shows how galaxy occupation favors different concentrations in each mass bin. The mass bins are chosen to contain equal number of galaxies. The $x$-axis shows the rank of halo concentration within a small halo mass bin, normalized to within $-0.5$ and 0.5. By definition, in each mass bin, $50\%$ of halos will have a positive $c$ rank whereas the other $50\%$ will have a negative $c$ rank. The black histogram shows the distribution of all halos. The orange histogram shows the distribution of the halo concentration of the central galaxies, whereas the green curve shows the distribution of concentration of the satellite galaxies. All the distributions are normalized to 1, so the relative magnitude of the different histograms is not meaningful. The dashed lines show the mean of the distributions. If there is no galaxy secondary bias, i.e. the mass-only assumption is true, then we expect the orange and green histograms to both resemble the black histogram, and the three dashed lines to overlap at 0. 

Within our DESI LRG mock, the centrals show a strong preference for the more concentrated halos per mass bin, as the orange histogram significantly tilts towards positive $c$ rank, with the mean concentration rank  consistently larger than 0. This preference also appears to be mass-dependent, with galaxies in less massive halos showing a stronger preference. This makes sense since we expect LRG centrals to trace older halos. For the most massive halos, this preference becomes less relevant as almost all halos have one central LRG regardless of secondary properties. For the satellites, the results are much noisier due to the limited sample size. However, we see a consistent preference for less concentrated halos among the satellites. This makes sense as the less concentrated halos are likely younger and had less time to undergo central mergers and tidal disruptions, which tend to disrupt/destroy satellites. The preference among the satellites appears less mass dependent, but this will need to be confirmed with a larger sample size.

\begin{figure}
    \centering
    \hspace*{-0.6cm}
    \includegraphics[width = 3.4in]{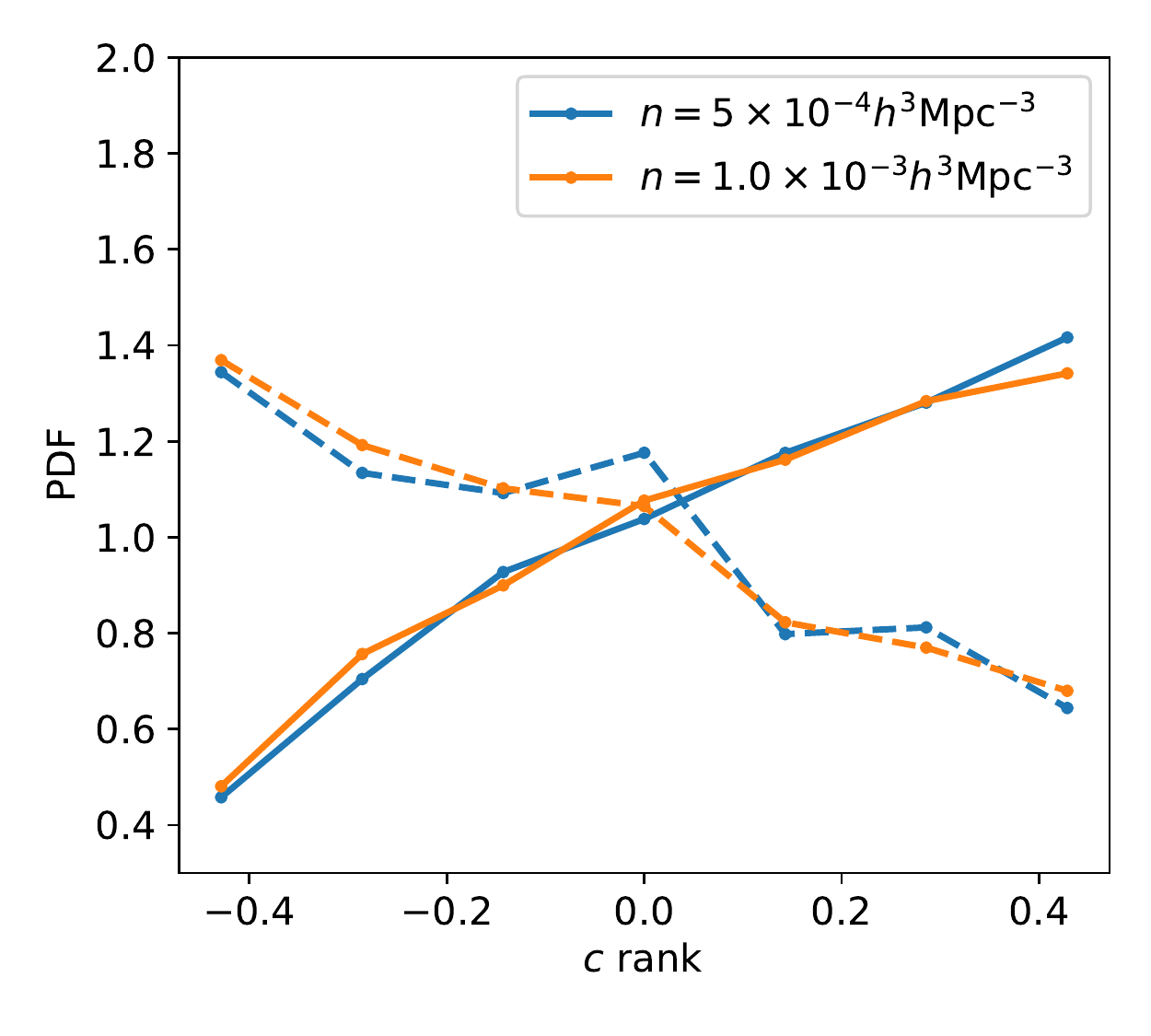}
    \vspace{-0.3cm}
    \caption{The distribution of DESI LRG mocks in the concentration rank combining all mass bins, comparing between two different number densities. The blue lines correspond to our DESI mock sample, while the orange corresponds to a number density twice as large. The solid lines show the distribution of the centrals, and the dashed lines show the distribution of the satellites. Overall, we see identical trends between the two number densities, showcasing clear evidence for concentrated-based secondary bias. }
    \label{fig:lrg_c_compare}
\end{figure}

The target density of $5\times 10^{-4}h^{3}$Mpc$^{-3}$ is a conservative estimate of the eventual density achieved by DESI. To test the robustness of our results against expected number density, we compare to the distribution of the halo concentration in a galaxy sample with twice the number density in Figure~\ref{fig:lrg_c_compare}. The centrals are shown in solid while the satellites are shown in dashed. Specifically, the blue represents the same sample that is plotted in Figure~\ref{fig:lrg_crank}, but combining all three mass bins. The orange curves correspond to a qualitatively similar galaxy sample, selected also with the color selection cuts quoted in Equation~\ref{equ:lrg_cuts}, but with a different magnitude correction that generates a larger number density. Overall, we find strong evidence for concentration-based bias, consistent across both samples. These concentration dependencies are also identified in the CMASS LRG clustering analysis carried out in \citet{2021bYuan}, where we also find centrals to prefer more concentrated halos whereas the satellites prefer the less concentrated halos.

\begin{figure*}
    \centering
    \hspace*{-0.3cm}
    \includegraphics[width = 7.1in]{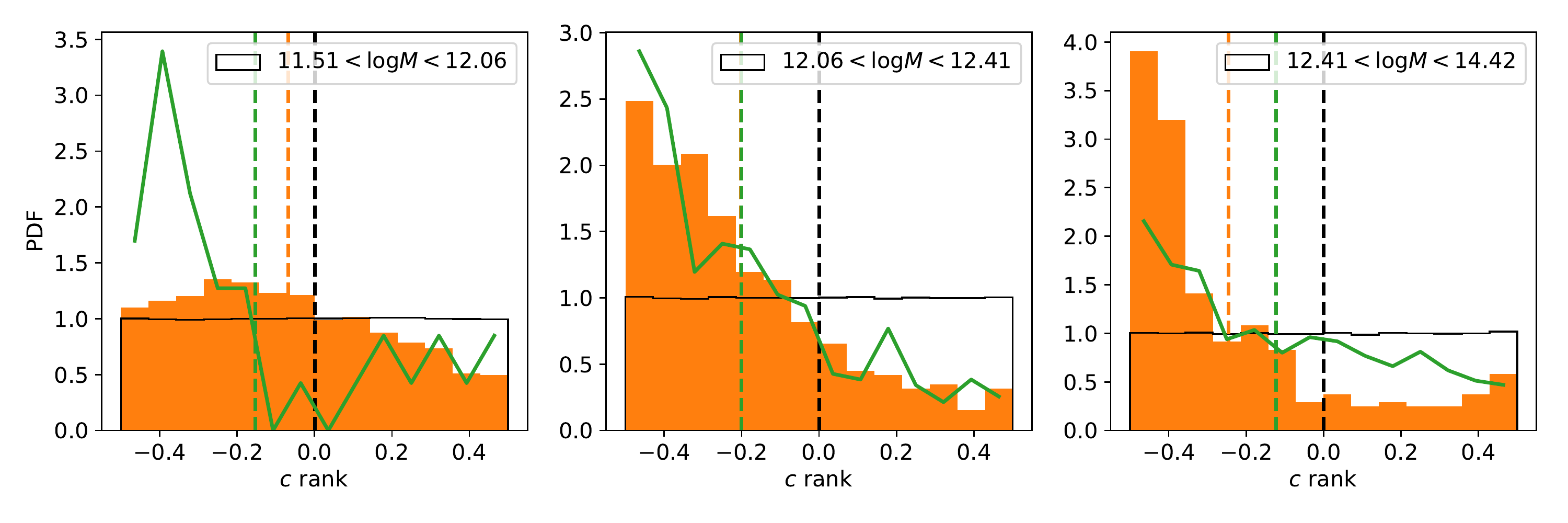}
    \vspace{-0.3cm}
    \caption{The halo occupation of DESI ELG mock galaxies as a function of concentration per mass bin. The $x$-axis shows the rank of halo concentration per mass bin, normalized to within $-0.5$ and 0.5. By definition, in each mass bin, $50\%$ of halos will have a positive $c$ rank whereas the other $50\%$ will have a negative $c$ rank. The black histogram shows the distribution of halos within the mass bin, showing a largely symmetric distribution around 0. The orange histogram shows the distribution of the halo concentration of the central galaxies, whereas the green curve shows the distribution of the satellite galaxies. all distributions are normalized to 1, so the relative magnitude of the distributions are not meaningful. The dashed lines represent the mean of the distributions. We see strong preference among the ELGs, with both centrals and satellites preferring less concentrated halos. }
    \label{fig:elg_crank}
\end{figure*}

Figure~\ref{fig:elg_crank} shows the concentration-based galaxy secondary bias signature of the DESI ELG mock galaxies. Again, we have divided the galaxies into 3 mass bins, and show the distribution of centrals in the orange histogram and satellites in the green curve. We see strong dependencies on concentration, with both the centrals and satellites preferring less concentrated halos. This makes sense as we expect ELGs to occupy younger star-forming halos. For the centrals, interestingly, we also see a strong mass dependency, where ELGs in more massive halos show significantly stronger concentration dependency. We speculate that this is due to young massive halos tend to exhibit particularly energetic star formation due to recent mergers. The ELG satellites also strongly prefer less concentrated halos, but there does not appear to be significant mass dependency. 

\begin{figure}
    \centering
    \hspace*{-0.6cm}
    \includegraphics[width = 3.4in]{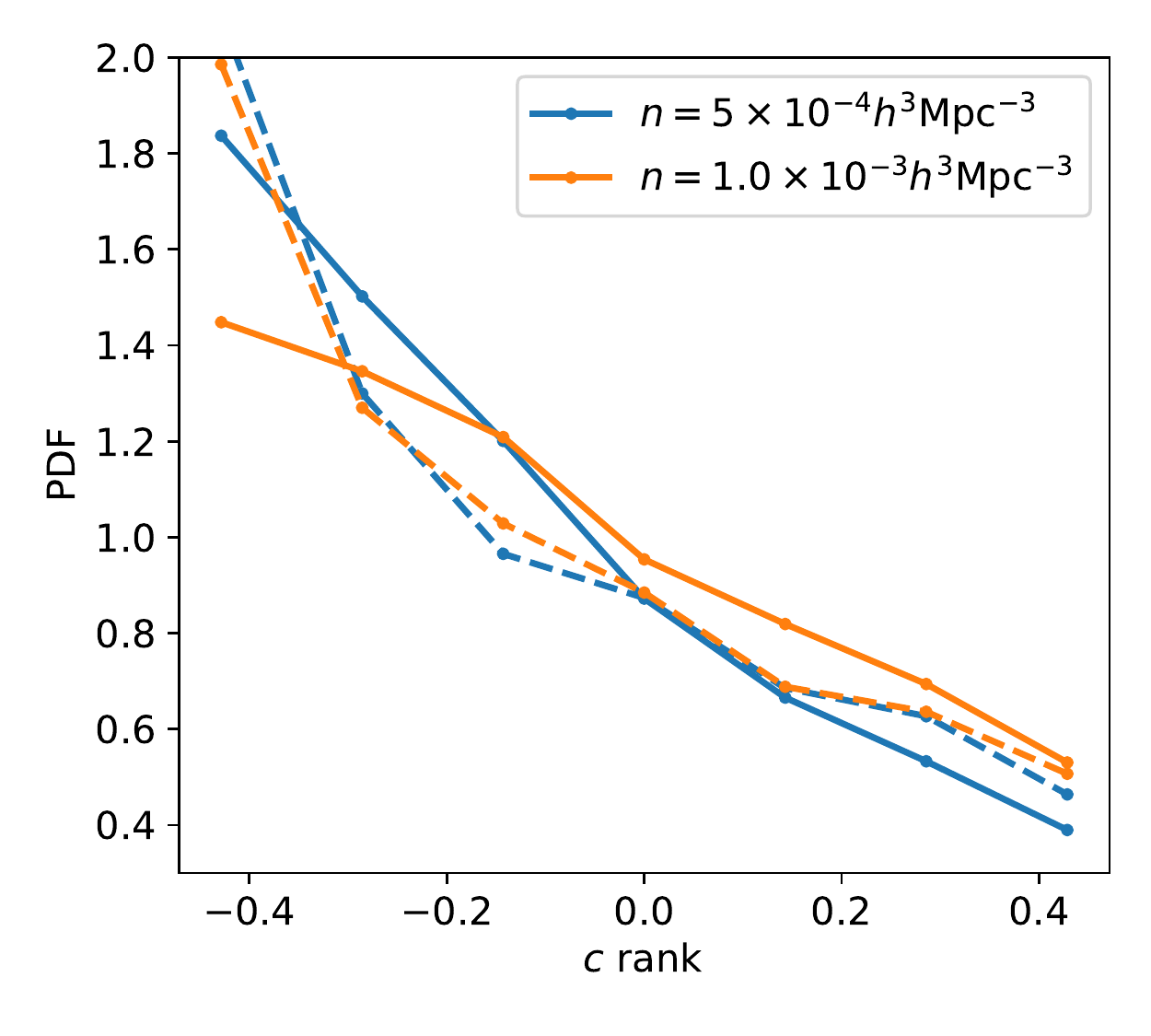}
    \vspace{-0.3cm}
    \caption{The distribution of DESI ELG mocks in the concentration rank combining all mass bins, comparing between two different number densities. The blue lines correspond to our DESI mock sample, while the orange corresponds to a number density twice as large. The solid lines show the distribution of the centrals, while the dashed lines show the distribution of the satellites. We find consistent trends across both number densities, with the lower density potentially showing stronger secondary bias among the centrals.}
    \label{fig:elg_c_compare}
\end{figure}

Figure~\ref{fig:elg_c_compare} showcases the halo concentration distribution of the mock ELG samples combining all mass bins, at two different number density, with DESI number density in blue, and a higher density in orange. The higher density mock ELG sample is again selected using Equation~\ref{equ:elg_cuts}, but with a different magnitude correction. Both samples show consistent preference for lower concentration halos, in both the centrals and satellites. 


To summarize this subsection, we find significant evidence that both the DESI LRG mocks and ELG mocks exhibit strong concentration-based secondary biases in their halo occupations. For the LRGs, we find that the centrals prefer more concentrated halos whereas the satellites prefer less concentrated halos. For the ELGs, we find that both the centrals and satellites prefer less concentrated halos. These findings are consistent with our expectations, and suggest that concentration-based assembly bias is likely an important ingredient in DESI HOD analyses. 

\subsubsection{Environment-based secondary bias}
\label{subsec:fenv}

In \citet{2020Hadzhiyska} and \citet{2021bYuan}, we find strong evidence that the halo local environment is a strong tracer of galaxy secondary bias, in both hydrodynamical simulations and observations. More recently \citet{2021Xu} and \citet{2021Delgado} used random forests to systematically identify the most important halo properties in an HOD framework, using hydrodynamical simulations and semi-analytic models, respectively. Both studies again found halo mass and halo environment to be by far the two most important galaxy occupation dependencies. All these studies combine to show that the local environment is a highly effective tracer of galaxy secondary bias, at least in terms of accurately predicting galaxy clustering down to 1-halo scales. In this subsection, we further substantiate that finding by directly measuring how galaxy occupation in our LRG and ELG mocks depends on the local environment in \textsc{IllustisTNG}. We also address the clustering predictions in Section~\ref{subsec:clustering}.

First, we follow \citet{2021bYuan} and define the local environment as the overdensity of neighboring subhalos within $r_{\mathrm{max}} = 5h^{-1}$Mpc but beyond the halo radius $r_{200c}$. Specifically, we can write down the environment definition as
\begin{equation}
    f_\mathrm{env} = \frac{M(r_{200c} < r < 5h^{-1}\mathrm{Mpc})}{\langle M(r_{200c} < r < 5h^{-1}\mathrm{Mpc})\rangle} - 1.
    \label{equ:fenv}
\end{equation}
Note that this definition is different in detail to that used in \citet{2021bYuan}, where we used the $r_{98}$ as the halo radius definition. $r_{98}$ refers to the radius that encloses 98$\%$ of the halo mass. However, \citet{2021bYuan} used a different simulation suite with different measured halo properties. Thus, we use $r_{200c}$ here as $r_{98}$ is not readily available, but the two halo radius definitions should be qualitatively similar and we do not expect this change in definition to lead to any significant change in our results. Similar to concentration, we then rank the $f_\mathrm{env}$ among halos within the same mass bin, and then normalize the ranks to within $-0.5$ and 0.5. The choice of $r_{\mathrm{max}} = 5h^{-1}$Mpc is found to best capture the secondary bias signature in simulations and it also yields the best fit to data \citep[][]{2021Yuan}.

\begin{figure*}
    \centering
    \hspace*{-0.3cm}
    \includegraphics[width = 7.1in]{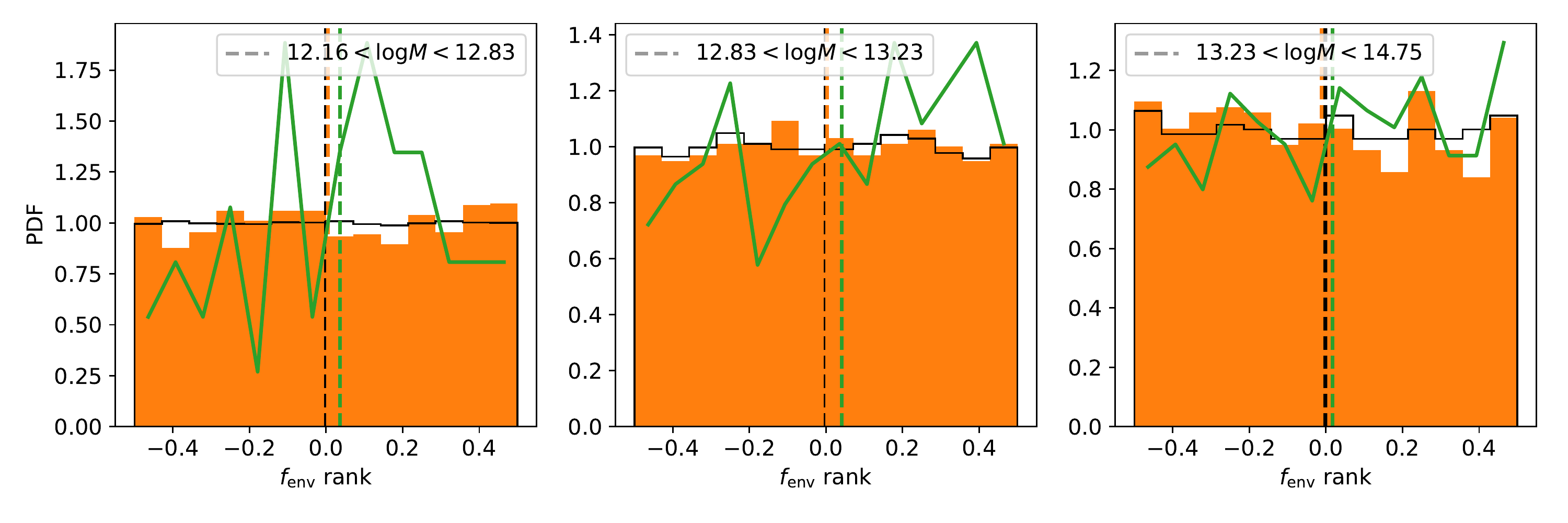}
    \vspace{-0.3cm}
    \caption{The halo occupation of DESI LRG mock galaxies as a function of local environment per mass bin. The $x$-axis shows the rank of halo environment within a small halo mass bin, normalized to within $-0.5$ and 0.5. By definition, in each mass bin, $50\%$ of halos will have a positive $f_{\mathrm{env}}$ rank whereas the other $50\%$ will have a negative $f_{\mathrm{env}}$ rank. The black histogram shows the distribution of halos within the mass bin, showing a largely symmetric distribution around 0. The orange histogram shows the distribution of the local environment of the central galaxies, whereas the green curve shows the distribution of the environment of the satellite galaxies. The dashed lines represent the mean of the histograms. There is little evidence for environment dependence in central galaxies, but there is clear evidence that satellite galaxies prefer halos in denser environment.}
    \label{fig:lrg_fenvrank}
\end{figure*}

\begin{figure}
    \centering
    \hspace*{-0.6cm}
    \includegraphics[width = 3.4in]{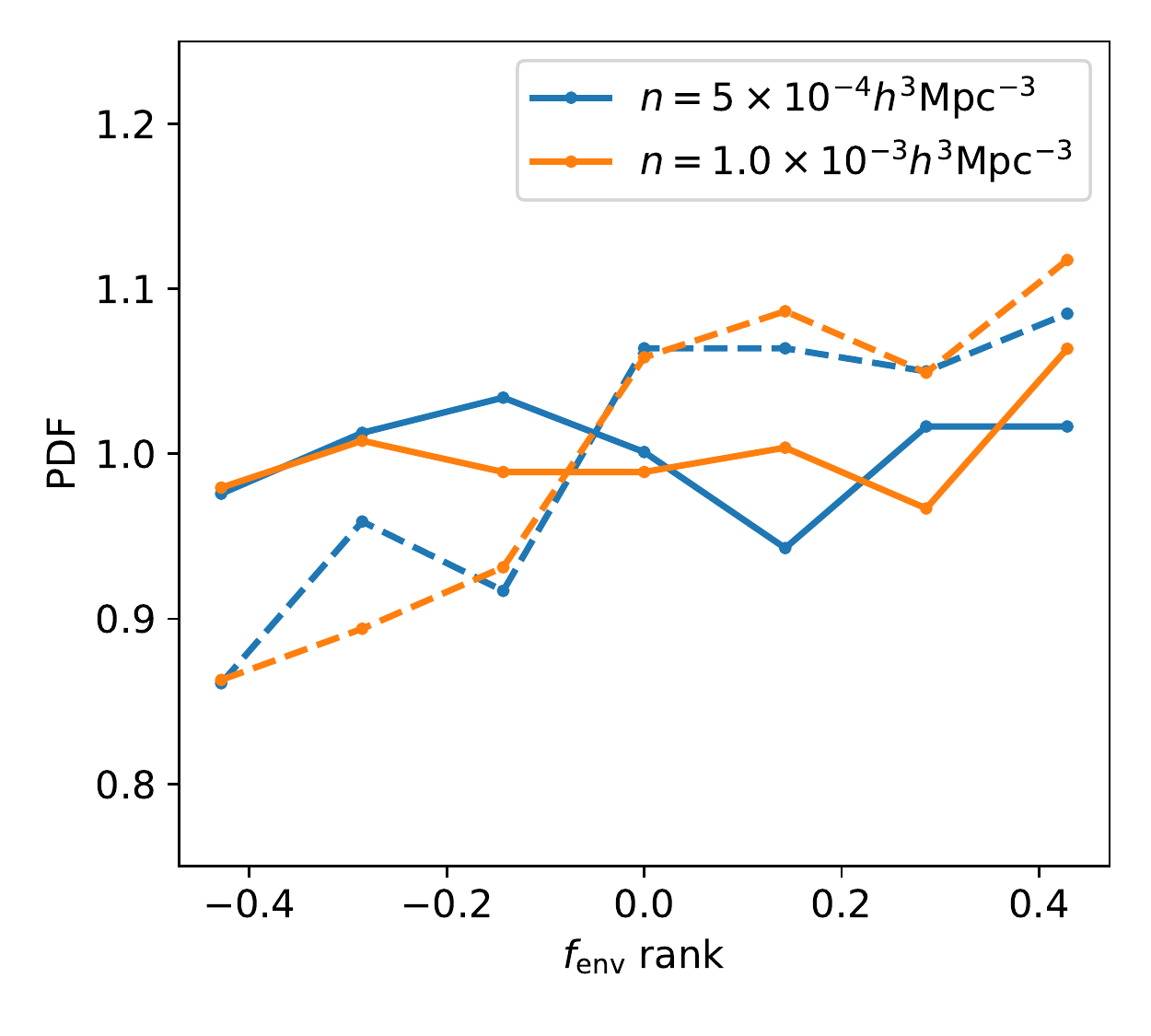}
    \vspace{-0.3cm}
    \caption{The distribution of DESI LRG mocks in the environment rank combining all mass bins, compared between two different number densities. The blue lines correspond to our DESI mock sample, while the orange corresponds to a number density twice as large. The solid lines show the distribution of the centrals, while the dashed lines show the distribution of the satellites. We do not see clear evidence for environmental dependence among the centrals, but we do see clear preference for denser environment among the satellites.}
    \label{fig:lrg_fenv_compare}
\end{figure}

Figure~\ref{fig:lrg_fenvrank} showcases the environment-based galaxy secondary bias signature of DESI LRG mock galaxies. Again the orange histogram shows the distribution of centrals while the green curve shows the distribution of the satellites. It is clear that while the centrals have little environmental preference, the satellites appear to prefer halos in denser environments across all mass ranges. Figure~\ref{fig:lrg_fenv_compare} highlights the environmental dependence by stacking all three mass bins, and comparing to a higher density LRG sample. Across both number densities, we see that the satellites prefer halos in denser environments while central distributions remain mostly flat. 

This finding is at least partially consistent with the observational constraints of \citet{2021bYuan}, where we find weak evidence for concentration-based secondary bias among BOSS LRGs but strong evidence for environment-based secondary bias. We defer a more detailed discussion to Section~\ref{subsec:prev_studies}. It is also reassuring that the dependence we see on environment rank is consistent with a linear model, as we adopt such a linear model for secondary biases in our HOD analyses in \citet{2021Yuan, 2021bYuan}.

\begin{figure*}
    \centering
    \hspace*{-0.3cm}
    \includegraphics[width = 7.1in]{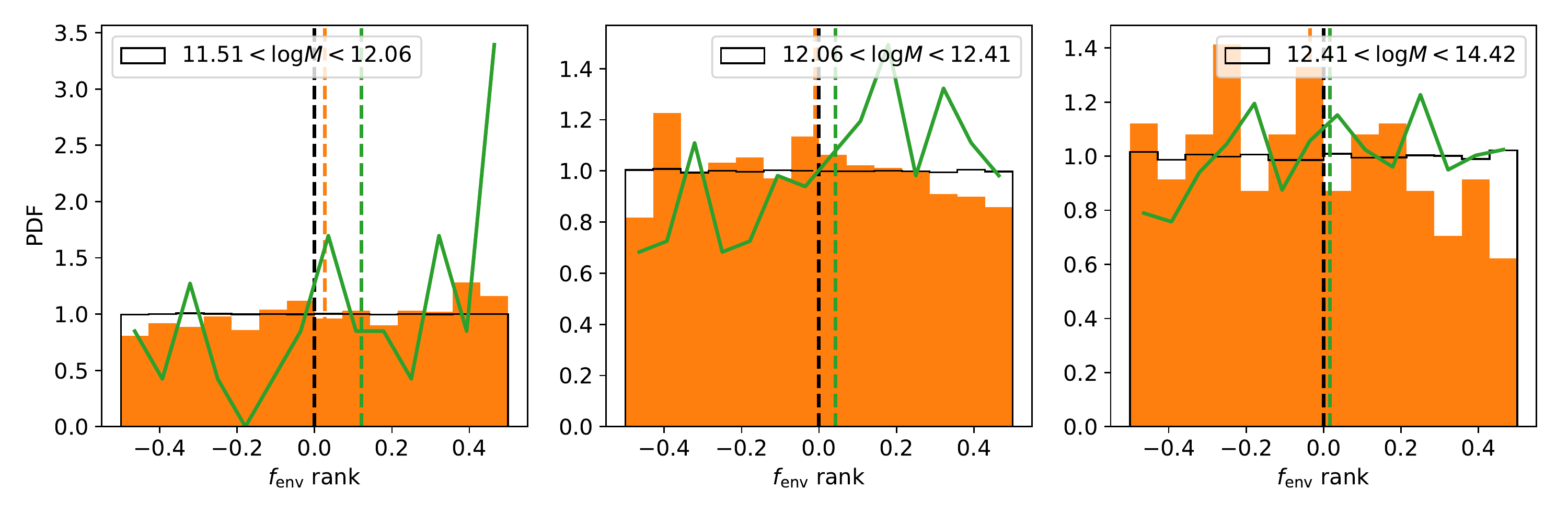}
    \vspace{-0.3cm}
    \caption{The halo occupation of DESI ELG mock galaxies as a function of local environment per mass bin. The $x$-axis shows the rank of halo environment within a small halo mass bin, normalized to within $-0.5$ and 0.5. The black histogram shows the distribution of halos within the mass bin, showing a largely symmetric distribution around 0. The orange histogram shows the distribution of the environment of the central galaxies, whereas the green curves show the distribution of the satellites. The dashed lines represent the mean of the distributions. We see clearly that the satellites have a clear preference for halos in denser environments across all mass bins. The centrals have a slight preference for halo in denser environments at low halo masses, but seems to reverse to preferring halos in less dense regions at higher halo mass.}
    \label{fig:elg_fenvrank}
\end{figure*}

Figure~\ref{fig:elg_fenvrank} showcases the environment-based galaxy secondary bias signature of DESI ELG mock galaxies. Similar to what we find for LRGs, there is a clear preference for halos in denser environments among the satellite galaxies, across all mass bins. This trend is potentially mass dependent, with the preference becoming weaker at higher halo masses. For the central galaxies, there appears to be a preference for halos in denser environments at low halo masses, but at higher masses, the secondary bias signature changes sign and shows a slight preference for halos in less dense environments. However, the signature for centrals is rather weak and should not be over-interpreted. 

\begin{figure}
    \centering
    \hspace*{-0.6cm}
    \includegraphics[width = 3.4in]{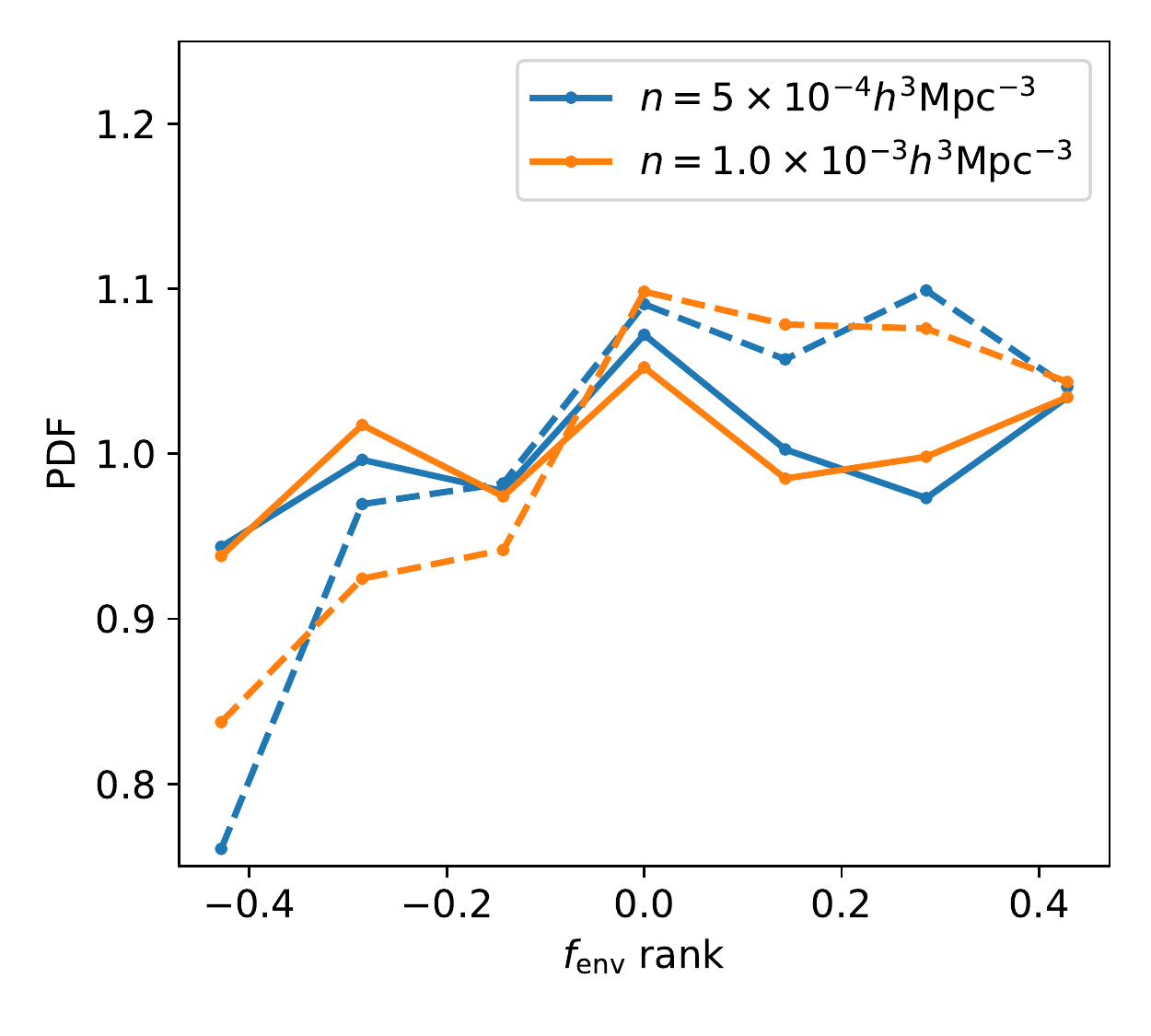}
    \vspace{-0.3cm}
    \caption{The distribution of DESI ELG mocks in the environment rank combining all mass bins, compared between two different number densities. The blue lines correspond to our DESI mock sample, while the orange corresponds to a number density twice as large. The solid lines show the distribution of the centrals, while the dashed lines show the distribution of the satellites. We again see clear preference for denser environment among the satellites. There also appears to be a slight preference for denser environments among the centrals. These trends seem to become stronger in the denser sample. }
    \label{fig:elg_fenv_compare}
\end{figure}

Figure~\ref{fig:elg_fenv_compare} highlights the environmental dependence by showcasing the stacked signal across all three mass bins, with centrals shown in solid lines and satellites shown in dashed lines. We compare two different number densities, DESI density in blue, and a higher density sample in orange. The satellites have a strong preference for denser environments, while the trend for centrals is much less significant, at least at DESI number density. However, at double the number density, the trend for both centrals and satellites appear to be stronger. This suggests that, at least for ELGs, the environmental secondary bias is more significant for lower mass higher density populations.

\section{Discussions}
\label{sec:discuss}
\subsection{Results from a mass-selected LRG mock sample}
A common way of selecting the LRG mock sample in studies utilizing hydrodynamical simulations and semi-analytic models is simply by selecting the most massive stellar objects in the simulation until we reach the desired number density. This scheme is widely assumed to result in a galaxy sample that resembles an LRG sample. In this section, we test such assumptions by comparing a mass-selected LRG sample to our color-magnitude selected sample, when matched to the same number density. Referring back to Figure~\ref{fig:lrg_sample} for some basic intuition, we see a mass-selected sample can deviate from the color-selected sample by including more objects with higher star formation rate. 
At the DESI LRG number density of $5\times 10^{-4}h^{3}$Mpc$^{-3}$, we find the mass-selected sample recovers largely the same properties as the color selected sample. Specifically, we find radial bias favoring larger radii in low mass halos, again likely due to the over-linking in FOF halos. In terms of velocity bias, we find $\alpha_c\approx 0.15$ and $\alpha_s\approx 1$, with larger satellite velocity biases in more massive halos.

\begin{figure}
    \centering
    \hspace*{-0.6cm}
    \includegraphics[width = 3.4in]{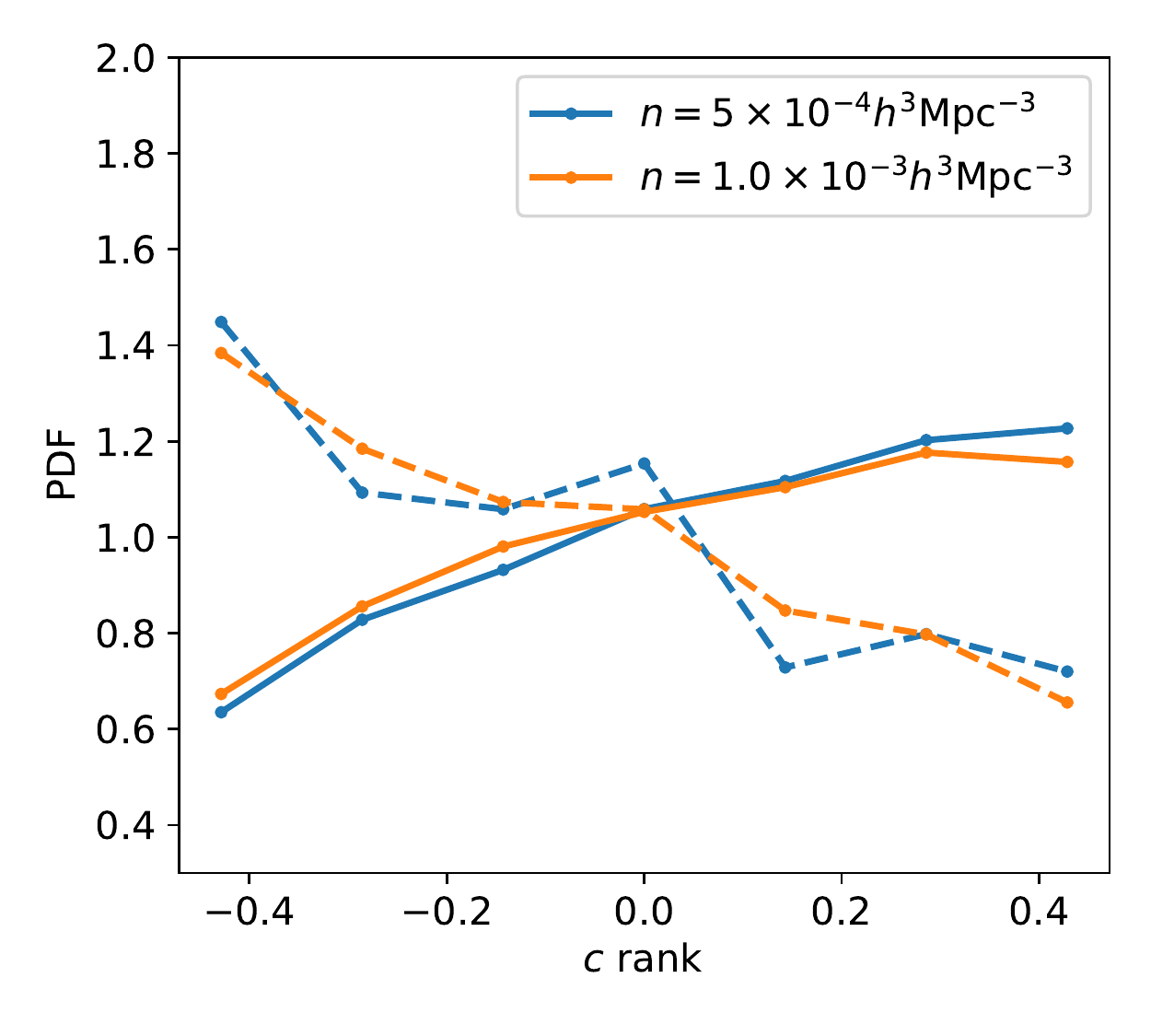}
    \vspace{-0.3cm}
    \caption{The distribution of mass-selected LRG mocks as a function of halo concentration, comparing between two different number densities. The blue lines correspond to our DESI mock sample, while the orange corresponds to a number density twice as large. The solid lines show the distribution of the centrals, and the dashed lines show the distribution of the satellites. We find the same qualitative trends as the color-selected sample, but the secondary bias among the centrals appear to be weaker. }
    \label{fig:lrg_c_compare_masscut}
\end{figure}

While we recover the same qualitative trends as the color-selected sample, we do see minor differences in the secondary bias signatures. Figure~\ref{fig:lrg_c_compare_masscut} showcases the distribution of mass-selected LRGs as a function of halo concentration rank, comparing two different number densities. Compared to the color-selected sample shown in Figure~\ref{fig:lrg_c_compare}, we see that the centrals continue to show preference for higher halo concentration while the satellites favor the less concentrated halos. However, the strength of the central dependency appears to be weaker than in the color-selected LRG sample. This is consistent with the fact that the mass selection allows for more star-forming galaxies, which tend to reside in younger less concentrated halos. 

\begin{figure}
    \centering
    \hspace*{-0.6cm}
    \includegraphics[width = 3.4in]{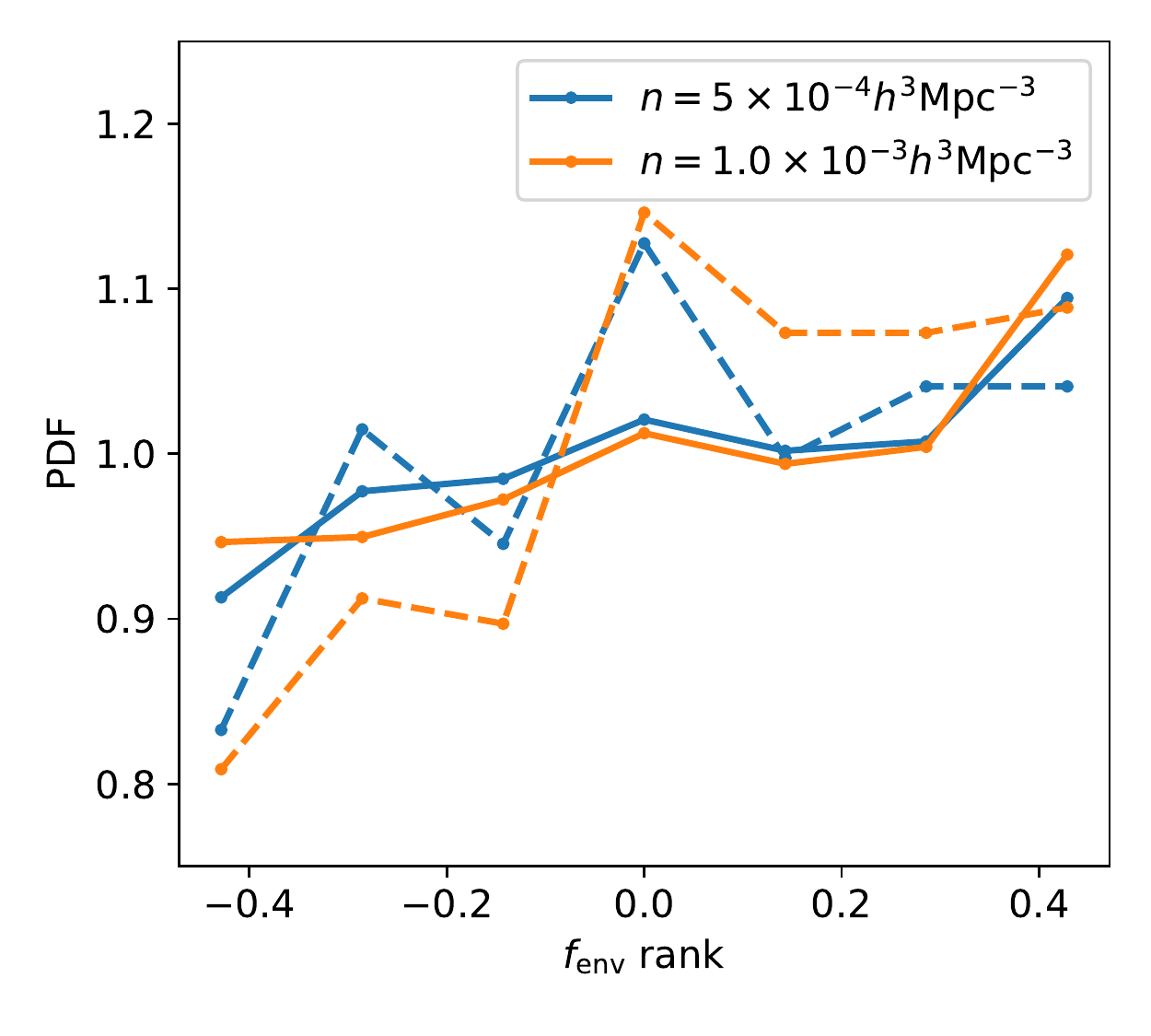}
    \vspace{-0.3cm}
    \caption{The distribution of mass-selected LRG mocks as a function of local environment, comparing between two different number densities. The blue lines correspond to our DESI mock sample, while the orange corresponds to a number density twice as large. The solid lines show the distribution of the centrals, and the dashed lines show the distribution of the satellites. The centrals and the satellites both seem to favor halos in denser environments.}
    \label{fig:lrg_fenv_compare_masscut}
\end{figure}

In terms of secondary bias as a function of local environment, the mass-selected sample also shows some interesting differences compared to the color-selected samples. Figure~\ref{fig:lrg_fenv_compare_masscut} shows the environment dependence of the mass-selected LRG sample. Compared to the color-selected sample shown in Figure~\ref{fig:lrg_fenv_compare}, the main difference is that the mass-selected centrals show a clear preference for halos in denser environments, contrary to the lack of preference found in the color-selected sample. This could be again related to the inclusion of some higher star-formation rate galaxies, as halos in denser environments tend to undergo mergers and exhibit higher accretion rates, thus contributing to stronger star formation \citep[e.g.][]{2009Fakhouri, 2010Genel, 2017Romano}. 

For the mass-selected sample, we continue to find the same clustering signature as the color-selected sample. We do not repeat those plots for brevity. Overall, we find the mass-selected sample to be qualitatively consistent with the color-selected sample, while there are unexpected differences in secondary biases that should serve as important caveats for future simulation-based mock galaxy studies. 

\subsection{Comparison to previous studies}
\label{subsec:prev_studies}
Several previous studies have also tested aspects of the HOD framework using \textsc{IllustrisTNG}. \citet{2019Bose} studied the galaxy-halo connections of mass-selected mock galaxy samples of number density $0.032 h^3$Mpc$^{-3}$ and $0.016 h^3$Mpc$^{-3}$, which is about 30-60 times denser than our DESI LRG and ELG mock samples, resulting in significantly lower halo mass and lower bias. In Figure~11 of \citet{2019Bose}, the authors compared the radial distribution of the satellite galaxies to that of the dark matter profile. The authors found that the satellite profile is largely consistent with the dark matter profile, except in the lowest mass bins, where the satellites appear more concentrated towards the halo core. This seemingly contradicts our findings in Figure~\ref{fig:lrg_radial} and Figure~\ref{fig:elg_radial}. This is due to the large difference in galaxy bias between our samples and the samples used in \citet{2019Bose}. In fact, if we use a mass-selected sample matching the much larger number density of \citet{2019Bose}, we similarly find that the radial profile of satellites are largely consistent with that of the dark matter. And if we use a mass-selected sample matching the DESI LRG density, then we again find that the satellite profile is significantly broader than the halo profile in the lower mass bins. This suggests that different galaxy populations can have significantly different radial distributions, especially if the different populations occupy different regions of the specific SFR vs stellar mass plane. \citet{2019Bose} also finds the local environment to have a rather mild impact on galaxy occupation, compared to other secondary bias indicators such as formation redshift and halo spin. This is consistent with what we are finding in our samples, despite the major differences between the samples. Their work, however, does not explore the effect of these secondary biases on the galaxy clustering. 

\citet{2021bHadzhiyska} employed a similar color-selection technique and extracted an DESI-like mock ELG sample, but it did not use the magnitude correction to match the DESI number density, resulting in an over-estimated ELG number density of $1\times 10^{-3}h^3$Mpc$^{-3}$. That study found a baseline HOD (Figure~4) that is consistent with ours (Figure~\ref{fig:elg_hod}). In Figure~6 of \citet{2021bHadzhiyska}, the authors examined the impact of local environment on ELG occupation by contrasting the HOD of the ELGs in the top and bottom $20\%$ environment. The authors found modest evidence for environment-based secondary bias in the color-selected sample and significantly stronger evidence in the mass-selected sample. Their findings are consistent with our results (summarized in Figure~\ref{fig:elg_fenvrank}), where we find a mild preference for centrals to occupy halos in less dense regions at higher mass, whereas the satellites have a mild preference for halos in denser regions, particularly at lower halo mass. It is worth noting that the two analyses also use two different definitions of local environment. \citet{2021bHadzhiyska} calculates local density from a overall smoothed density field, whereas we explicitly compute dark matter density, excluding the host halo.

\subsection{Clustering Analysis}
\label{subsec:clustering}
In this section, we extend the primary analysis of this paper by examining the clustering signatures of the mock LRG/ELG samples, particularly the contributions due to secondary biases. However, we preface this section by cautioning the readers that our clustering analysis suffers from poor statistical significance due to the limited simulation volume. We reserve a more rigorous clustering analysis for a later study, when larger hydrodynamical simulations become available. 

Broadly speaking, the impact on galaxy clustering in the 2-halo regime is a combination of variations in occupation (galaxy secondary bias) and how halo clustering depends on secondary halo properties, an effect known as halo assembly bias \citep[e.g.][]{2007Croton, 2018Mao}. In principle, the galaxy secondary bias can simply be treated as a reweighting of the halos in the clustering calculation. In this section, we focus on the clustering signatures of galaxy secondary bias by disentangling its effects from that of halo assembly bias through a resampling routine.

Specifically, to fully remove the effects of secondary biases given a galaxy sample, we divide the sample into a set of narrow halo mass bins (measured on the DMO halos). Then we tabulate the number of centrals and satellites in each mass bin and compute the corresponding average number of centrals and satellites per halo. Finally, to create a galaxy mock without secondary bias, we paint centrals and satellites onto DMO halos as a function of halo mass using the tabulated values, assuming a Bernoulli distribution for the centrals and a Poisson distribution for the satellites. To determine the satellites' radial positions within each halo, we resample from the satellite radial distribution within that mass bin. By definition, this resampled set of centrals and satellites match the original HOD, but have no galaxy secondary biases. Thus, if we compare the clustering signature of the resampled galaxy sample to that of the original sample, the difference is purely due to the secondary biases. 

An important technicality is that the shuffling not only erases secondary biases in galaxy occupation, but it also removes any non-Poisson signatures in the satellite occupation, as we impose a Poisson satellite occupation in the shuffling. Thus, the difference between the mock sample and the shuffled sample also technically includes the clustering signatures due to non-Poisson satellite distribution. In Section~\ref{subsec:poisson}, we showed that the second moment of the satellite occupation is indeed consistent with a Poisson prediction for both samples, but we stopped short of actually proving that the satellite occupation is in fact Poissonian. When a larger mock sample becomes available and we find statistically significant evidence for non-Poisson satellite occupation, then we would need to devise a new shuffling scheme for the satellites. For now, we assume the satellites are indeed Poissonian.  

\begin{figure}
    \centering
    \hspace*{-0.6cm}
    \includegraphics[width = 3.4in]{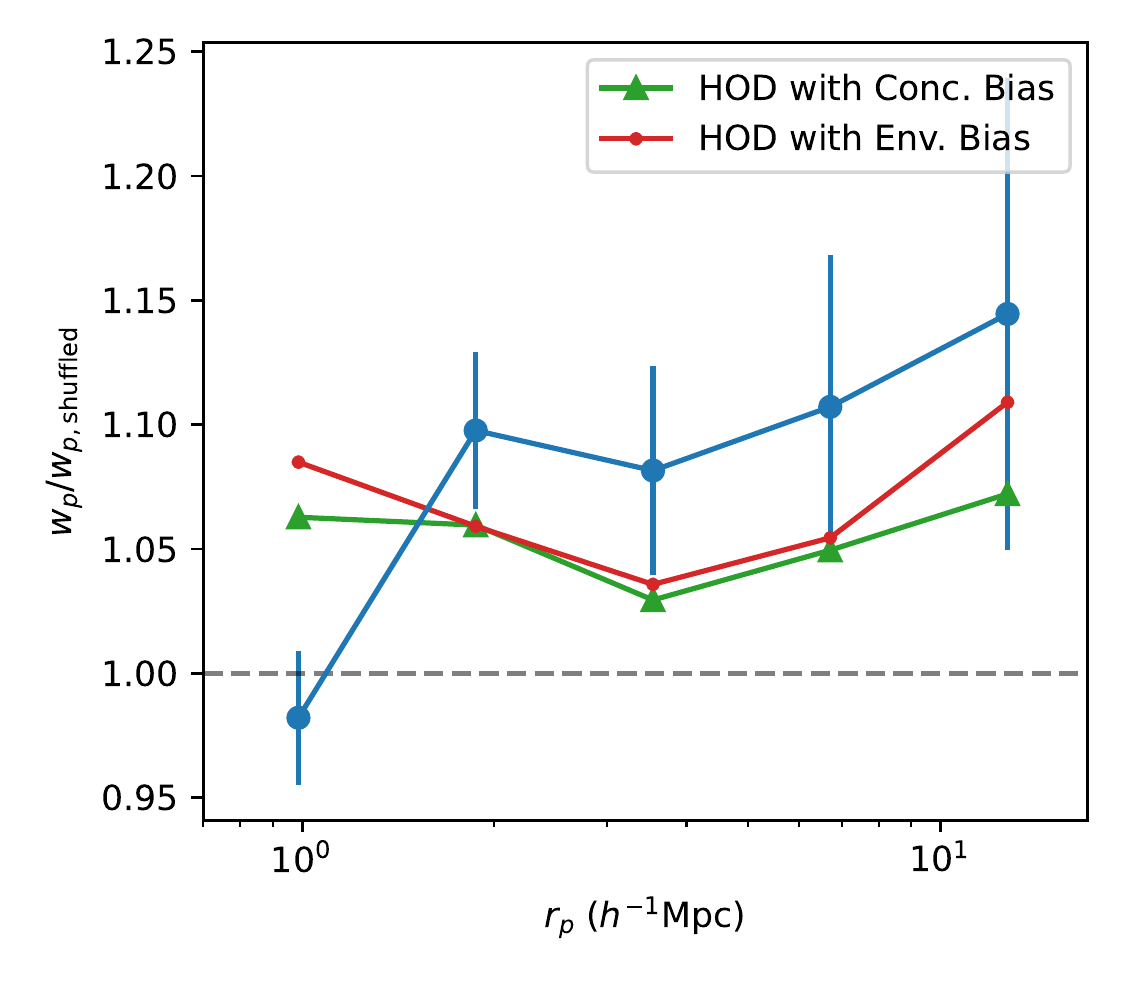}
    \vspace{-0.6cm}
    \caption{The clustering signature of galaxy secondary biases for the mock LRG sample. The blue line showcases the total excess in clustering measured between the LRG sample and its vanilla HOD counterpart. The error bars show the level of jackknife sample variance. The green and red curves show the amount of excess clustering induced by modifying the vanilla HOD with the same amount of secondary dependency (on concentration and environment respectively) as measured in the LRG mock. These two curves have the same level of sample variance as the blue curve.}
    \label{fig:gab_lrg}
\end{figure}

\begin{figure}
    \centering
    \hspace*{-0.6cm}
    \includegraphics[width = 3.4in]{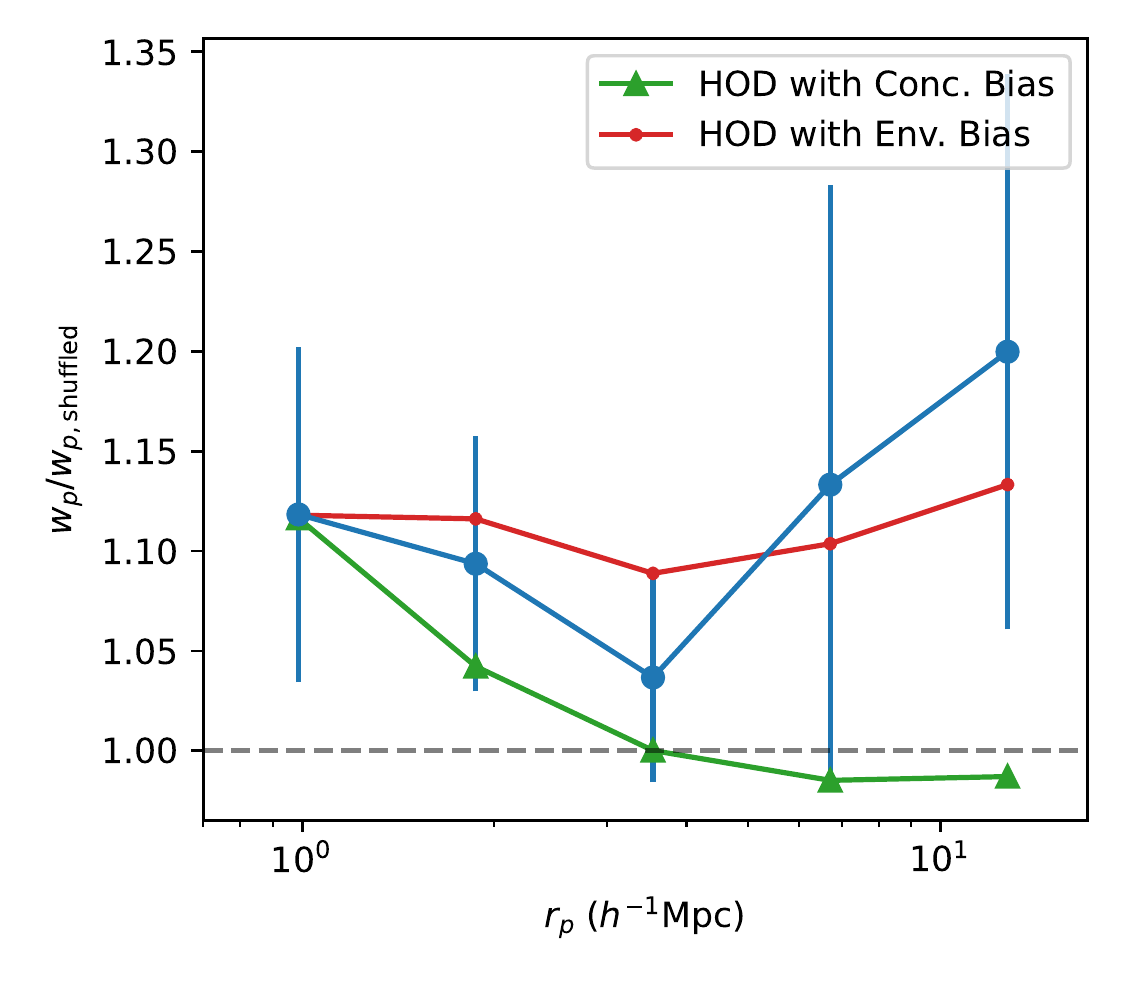}
    \vspace{-0.6cm}
    \caption{The clustering signature of galaxy secondary biases for the mock ELG sample. The blue curve shows the total excess clustering measured between the ELG mock and its vanilla HOD counterpart. The error bars show the jackknife sample variance. The green and red curves show the amount of excess clustering induced by modifying the vanilla HOD with the same amount of secondary dependency (on concentration and environment respectively) as measured in the ELG mock. These two curves have the same level of sample variance as the blue curve. }
    \label{fig:gab_elg}
\end{figure}

We measure the clustering signature due to galaxy secondary biases for both the mock LRG sample and mock ELG sample, as showcased by the blue curves in Figure~\ref{fig:gab_lrg} and Figure~\ref{fig:gab_elg}, respectively. We specifically showcase the projected 2-point correlation $w_p$, as a function of projected separation. The errorbars are generated from 27 jackknifes of the simulation volume. To maximize signal-to-noise, each clustering measurement is also repeated along all three axes then averaged. For the mock LRGs, we see a 10-15$\%$ excess in projected clustering in the original sample compared to the resampled LRG sample at scales $r_p > 1 h^{-1}$Mpc. The signature for LRGs has modest statistical significance and is qualitatively consistent with the findings of \citet{2020Hadzhiyska} and \citet{2020Xu}, though they find a somewhat larger signature, around $20\%$. However, their galaxy samples are mass-selected and matched to a much higher number density. We repeat our analysis for mock LRG samples with number density $n = 1.0\times 10^{-3}h^3$Mpc$^{-3}$ and $n = 1.5\times 10^{-3}h^3$Mpc$^{-3}$, and find the same clustering signature. 

For the ELG mock sample, we also find a 5-20$\%$ clustering signature due to galaxy secondary biases, though with very low statistical significance. This is likely due to the lower bias of the mock ELG sample, resulting in fewer pairs at these smaller scales. The slight positive signature is consistent with the findings of \citet{2021bHadzhiyska}, where the authors also found a slightly positive trend, that is also statistically insignificant despite using a higher density ELG sample.

Next we attempt to determine which secondary dependencies (concentration vs. environment) is responsible for the observed clustering signature. We adopt a 2-dimensional HOD (2D HOD) routine, where we tabulate the number of centrals and satellites in each 2D bin of halo mass and a secondary property. Then we populate the DMO halos with this tabulated 2D HOD. We determine the radial position of the satellites in the same fashion as for the mass-only case, where we re-sample from the satellite radial distribution within the 2D bin. The resulting mock has the same amount of dependency on the chosen secondary property as the original mock sample, but no dependency on any other properties. If this 2D HOD mock produces the same amount of excess clustering as the original sample, then we have preliminary evidence that this chosen secondary property is responsible for the clustering signature. We emphasize that the low signal-to-noise of our clustering measurements limits the level of confidence in such claims, but this remains an interesting exercise and will be significantly more valuable with a larger hydrodynamical simulation. 

Figure~\ref{fig:gab_lrg} showcases this analysis for the mock LRG sample. The blue curve represents the total amount of excess clustering in the LRG sample, as measured between the clustering of the LRG mock and its baseline HOD counterpart populated on the DMO halos. The green and red curves correspond to the excess clustering induced by introducing concentration-based and environment-based secondary biases in the baseline HOD mock, respectively. The amplitude of the secondary dependencies are calibrated to be the same as the full-physics LRG mock. Since the 2D HOD mock has the same sample size and similar clustering as the original mock sample, we expect the level of sample variance on these 2D HOD measurements to  be the same as the blue curve. This suggests, with weak statistical significance, that the full clustering signature can be explained by a combination of concentration-based and environment-based secondary dependencies. The discrepancy in the first bin is not meaningful, due to the radial re-sampling we apply to the satellite positions. 

Figure~\ref{fig:gab_elg} repeats this exercise for the mock ELG sample. Again the blue curve showcases he full excess clustering signature where the green and red curves show the excess clustering induced by applying secondary dependencies on to the baseline HOD mock. The amplitude of the secondary dependencies is again calibrated on the full-physics ELG mock. For this sample, we find that the environment-based secondary bias produces the majority of the excess clustering, whereas the concentration-based piece plays a smaller role. Again, these claims are statistically weak due to the large sample variance and the fact that it is unclear we see excess clustering in the blue curve in the first place.

While we cannot claim either secondary biases to be more important for clustering due to the limited signal-to-noise, we have shown that despite the environment dependence appearing weaker than the concentration dependence (e.g., comparing
Fig. 11 and 15), it is at least as important as concentration in their effects on clustering. However, other recent studies have found environment to be by far the more important secondary halo property in predicting small-scale clustering when looking at higher density samples in hydrodynamical simulations and semi-analytic galaxy catalogs. Specifically, \citet{2021Delgado} used random forests to systematically identify the most important halo properties for galaxy clustering in an HOD framework, on a $n = 1.4\times 10^{-3}h^3$Mpc$^{-3}$ LRG sample in \textsc{IllustrisTNG}. \citet{2021Xu} conducted a similar analysis on a much larger semi-analytic galaxy sample. Both studies found halo mass and halo environment to be by far the two most important galaxy occupation dependencies. This hierarchy of halo properties is also separately supported by analyses using N-body simulations (see Figure~1 of \citet{2021bYuan}), where we previously found that the clustering derivatives against environment-based bias is much stronger than derivatives against concentration-based bias. Nevertheless, the clustering exercise we demonstrated in this section is novel and interesting and should become highly informative when larger hydrodynmical volumes become available. 
Also, a greater volume will allow us to probe higher mass halos, where clustering becomes particularly sensitive to details and extensions of the HOD. 

\section{Conclusions}
\label{sec:conclusion}
In this paper, we apply DESI selection cuts to \textsc{IllustrisTNG} galaxies to build mock samples of DESI-like mock LRGs and ELGs. We study their galaxy-halo connections in the form of HODs and relevant HOD extensions. We summarize the key observations as the following:
\begin{itemize}
  \item The halo occupation of both the mock LRGs and ELGs appear well-described by their respective baseline HOD formulas (Figure~\ref{fig:lrg_hod} and Figure~\ref{fig:elg_hod}).
  \item The satellite occupation of both samples are consistent with a Poisson distribution (Figure~\ref{fig:lrg_poisson} and Figure~\ref{fig:elg_poisson}). 
  \item The satellite radial profiles of both samples show a bimodal distribution at low halo masses, speculatively due to halo finding issues, but we do not rule out physical explanations (Figure~\ref{fig:lrg_radial} and Figure~\ref{fig:elg_radial}). 
  \item We find strong evidence for central velocity bias in both samples, consistent with observational constraints. The satellites in both samples show a rather modest velocity bias, with an interesting mass dependency (Figure~\ref{fig:lrg_vel} and Figure~\ref{fig:elg_vel}). 
  \item In our investigation of galaxy assembly bias, we find strong concentration-based secondary bias. For LRGs, we find the centrals prefer older and more concentrated halos, whereas the satellites prefer younger and less concentrated halos (Figure~\ref{fig:lrg_crank}). Both ELG centrals and satellites prefer younger and less concentrated halos (Figure~\ref{fig:elg_crank}). 
  \item We find weaker but clear environment-based secondary biases among the satellites in both samples. In both samples, the satellites prefer halos in denser environments (Figure~\ref{fig:lrg_fenvrank} and Figure~\ref{fig:elg_fenvrank}). Additionally, the ELG centrals appear to prefer halos in less dense environments at higher halo mass. 
\end{itemize}
Additionally, we find our conclusions are robust against number density constraints, and we reproduce much of the same results when we adopt a stellar mass selection for the LRGs. 
We also conduct a preliminary clustering analysis where we found an excess clustering signature due to secondary biases in the LRGs. We do not find a statistically significant excess clustering signature for the ELGs. We also conduct a 2D HOD exercise to identify which secondary bias to be more important for clustering, while we show the environment-based bias to be at least as important as the concentration-based bias in clustering predictions, we do not reach any statistically significant conclusion as to which one is more important. However, other studies based on hydrodynamical simulations and semi-analytic studies do find environment to be the most important secondary halo property for clustering. 

In the broader context of cosmological analysis in the DESI era, this analysis serves several important purposes. First, it illustrates and informs the need for more sophisticated galaxy-halo connection models beyond vanilla HODs for DESI small-scale analysis, with the caveat that we do not claim that any specific hydrodynamical recipe gives us the full range of small-scale physics with high fidelity. Nevertheless, our analysis should inform small-scale analyses as to how to construct a realistic HOD model and what are the important extensions to include. Second, this study is an important step towards building more realistic mock galaxy catalogs for DESI and upcoming surveys by directly leveraging hydrodynamical simulations and relying less on analytic assumptions. Specifically, once we have summarized the galaxy-halo connection model in hydronamical simulations with a high-dimensional parametrized model, we can re-apply such model to a much larger N-body dark-matter-only simulation such as \textsc{AbacusSummit} \citep{2021Maksimova} to achieve both the fidelity of hydrodynamical simulations and the much greater volume offered by dark-matter-only simulations. All in all, this analysis should serve as a valuable reference for galaxy-halo connection modeling in upcoming clustering studies that aim to probe deeply non-linear scales. 

\section*{Acknowledgements}
We would like to thank Andrew Hearin, Lehman Garrison, Phil Mansfield, Risa Wechsler, Alexie Leauthaud, Joseph DeRose, Johannes Lange, Lars Hernquist for valuable discussions and feedback.

This work was supported by U.S. Department of Energy grant DE-SC0013718, NASA ROSES grant 12-EUCLID12-0004, NSF PHY-2019786, and the Simons Foundation.
SB is supported by the UK Research and Innovation (UKRI) Future Leaders Fellowship [grant number MR/V023381/1].

This work used resources of the National Energy Research Scientific Computing Center (NERSC), a U.S. Department of Energy Office of Science User Facility located at Lawrence Berkeley National Laboratory, operated under Contract No. DE-AC02-05CH11231.

\section*{Data Availability}

The simulation data are available at \url{https://www.tng-project.org/}. The generalized HOD framework referenced in this paper is available as a code package as a part of the \textsc{abacusutils} package at \url{http://https://github.com/abacusorg/abacusutils}.



\bibliographystyle{mnras}
\bibliography{biblio} 








\bsp	
\label{lastpage}
\end{document}